\documentclass[12pt]{article}
\usepackage{graphicx}
\usepackage{url,amstext,paralist}
\usepackage{hyperref}
\input{epsf}
\DeclareGraphicsExtensions{.pdf,.png}
\pdfoutput=1 

\def\tev{\textrm{TeV}}
\def\gev{\textrm{GeV}}
\def\pt{\ensuremath{p_{\rm T}}}
\def\met{\ensuremath{E_{\rm T}^{\rm miss}}}
\def\imb{\ensuremath{\textrm{$\mu$b}^{-1}}}
\def\ipb{\ensuremath{\textrm{pb}^{-1}}}
\def\ifb{\ensuremath{\textrm{fb}^{-1}}}
\def\to{\ensuremath{\rightarrow}}
\def\X{\ensuremath{\tilde\chi_1^0}}
\def\x{\ensuremath{\chi}}
\def\t#1{\tilde{ #1}}

\begin{document}

\title{ATLAS PHYSICS RESULTS}
\author{Vasiliki A.\ Mitsou\thanks{E-mail: vasiliki.mitsou@ific.uv.es}\\
{\small Instituto de F\'isica Corpuscular (IFIC), CSIC -- Universitat de Val\`encia,}\\ 
{\small Parc Cient\'ific de la U.V., C/ Catedr\'atico Jos\'e Beltr\'an 2,}\\
{\small E-46980 Paterna (Valencia), Spain}}
\date{\small November 14, 2013}
\maketitle
\begin{abstract}
The ATLAS experiment at the Large Hadron Collider at CERN has been successfully taking data since the end of 2009 in proton-proton collisions at centre-of-mass energies of 7 and 8~\tev, and in heavy ion collisions. In these lectures, some of the most recent ATLAS results will be given on Standard Model measurements, the discovery of the Higgs boson, searches for supersymmetry and exotics and on heavy-ion results.
\end{abstract}

\section{Introduction}\label{sc:intro}

Particle Physics aims at explaining the known content and forces of the Universe at a fundamental level. This knowledge is summarised in the Standard Model (SM)~\cite{sm}: It includes 12~elementary matter particles, whereas their interactions are mediated via 5~gauge bosons, induced by requiring local gauge invariance.

The Standard Model provides an excellent description of collider experimental data so far. LEP, SLC, Tevatron, $B$-factories and LHC data show that SM describes physics at energies up to $\sqrt{s}\sim200~\gev$ with respect to QCD and hadronic structure, precision EW physics, top quark properties and flavour physics. Despite its success, the SM suffers from a number of shortcomings: How we solve the hierarchy problem in the electroweak symmetry breaking (Higgs mechanism)? How are the neutrino masses generated? Are the neutrinos Dirac or Majorana particles? Which are the new sources for $CP$ violation necessary to explain the observed matter-antimatter baryon asymmetry? Gravitation is the only known force not included in the SM; how can be incorporated? The strong and electroweak interactions are not unified to a single coupling; how can grand unification be achieved? What is the nature of the dark matter and dark energy components of the Universe? 

The Large Hadron Collider (LHC)~\cite{lhc} at CERN in Geneva is designed, constructed and operates precisely in order to address these open issues of the Standard Model. In particular, the ATLAS experiment~\cite{atlas-det} has been successfully taking data since the end of 2009 till the first months of 2013 in proton-proton collisions at centre-of-mass energies of 7~and 8~\tev, and in heavy-ion collisions. An overview of the physics results obtained so far is given here.

The ATLAS detector~\cite{atlas-det} consists of an inner tracking system (inner detector, or ID) surrounded by a superconducting solenoid providing a 2~T magnetic field, electromagnetic and hadronic calorimeters, and a muon spectrometer (MS) embedded in three large superconducting toroid magnets arranged in an eight-fold azimuthal coil symmetry around the calorimeters. The ID~\cite{id} consists of silicon pixel~\cite{pixel} and microstrip~\cite{sct} superconducting detectors, surrounded by a gaseous transition radiation tracker~\cite{trt}. The electromagnetic calorimeter is a lead/liquid-argon (LAr) detector~\cite{larg}. Hadron calorimetry is based on two different detector technologies, with scintillator tiles or LAr as active media, and with either steel, copper, or tungsten as the absorber material~\cite{larg,tilecal}. The MS comprises three layers of different types of muon chambers for triggering and for track measurements~\cite{muon}.

The structure of this paper is as follows. Section~\ref{sc:sm} provides an overview of the SM measurements carried out with ATLAS. Sections~\ref{sc:top} and~\ref{sc:bphys} highlight recent results on top quark and flavour physics, respectively. In Section~\ref{sc:hi}, the runs and latest observations with heavy-ion collisions are discussed. The Higgs boson discovery and studies aiming at determining its properties are given in Section~\ref{sc:higgs}. In Sections~\ref{sc:susy} and~\ref{sc:exotics}, the latest results in searches for supersymmetry and for other beyond-the-SM (BSM) theoretical scenarios, respectively, are presented. The paper concludes with a summary and an outlook in Section~\ref{sc:summary}.

\section{Standard Model measurements}\label{sc:sm}

The measurement of the high-mass Drell-Yan differential cross section in proton-proton collisions at a centre-of-mass energy of 7~\tev\ at the LHC is reported in Ref.~\cite{dy}. Based on an integrated luminosity of 4.9~\ifb, the differential cross-section in the $Z\to e^+e^-$ channel is measured with the ATLAS detector as a function of the invariant mass, $m_{ee}$, in the range $116 < m_{ee} < 1500~\gev$, for a fiducial region in which both the electron and the positron have transverse momentum $\pt > 25~\gev$ and pseudorapidity $|\eta| < 2.5$. A comparison is made to various event generators and to the predictions of perturbative QCD calculations at next-to-next-to-leading order.

\begin{figure}[htb]
{\includegraphics[width=0.65\textwidth]{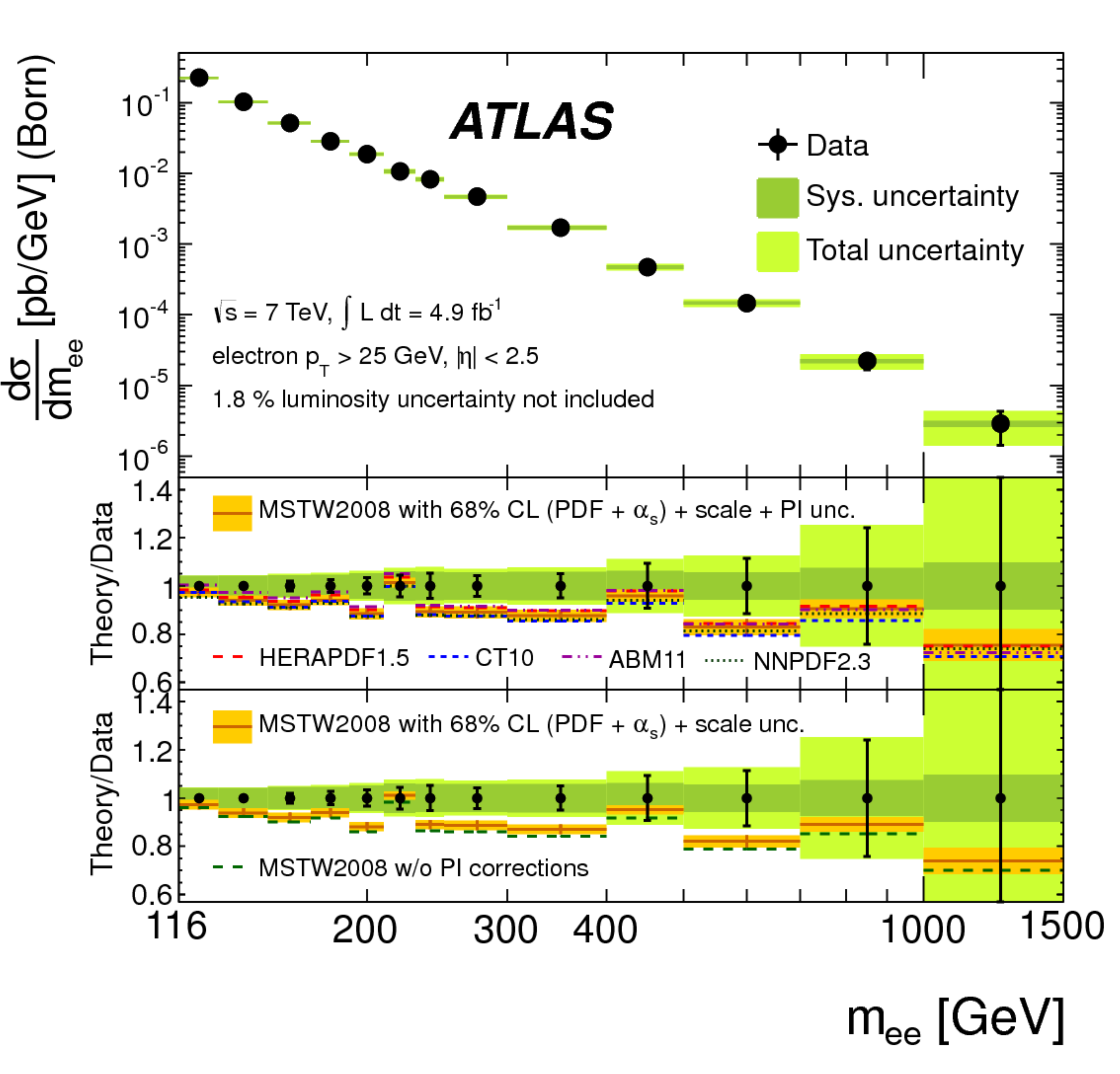}}
\caption{The measured differential cross section at the Born level within the fiducial region with statistical, systematic, and combined statistical and systematic (total) uncertainties. From Ref.~\cite{dy}. \label{fg:dy}}
\end{figure}

The measured differential cross section at the Born level within the fiducial region (electron $\pt > 25~\gev$ and $|\eta| < 2.5$) with statistical, systematic, and combined statistical and systematic (total) uncertainties, excluding the 1.8\% uncertainty on the luminosity is shown in Fig.~\ref{fg:dy}. The measurement is compared to FEWZ 3.1 calculations at NNLO QCD with NLO electroweak corrections using the $G_{\mu}$ electroweak parameter scheme. The predictions include an additional small correction from single-boson production in which the final-state charged lepton radiates a real $W$ or $Z$ boson. In the upper ratio plot, the photon-induced corrections have been added to the predictions obtained from the MSTW2008, HERAPDF1.5, CT10, ABM11 and NNPDF2.3 NNLO parton distribution functions (PDFs), and for the MSTW2008 prediction the total uncertainty band arising from the PDF, $\alpha_s$, renormalisation and factorisation scale, and photon-induced uncertainties is drawn. The lower ratio plot shows the influence of the photon-induced corrections on the MSTW2008 prediction, the uncertainty band including only the PDF, $\alpha_s$ and scale uncertainties. The comparison with event generators and PDFs show good agreement, therefore no sign of new physics is observed.

\begin{figure}[htb]
{\includegraphics[width=0.6\textwidth]{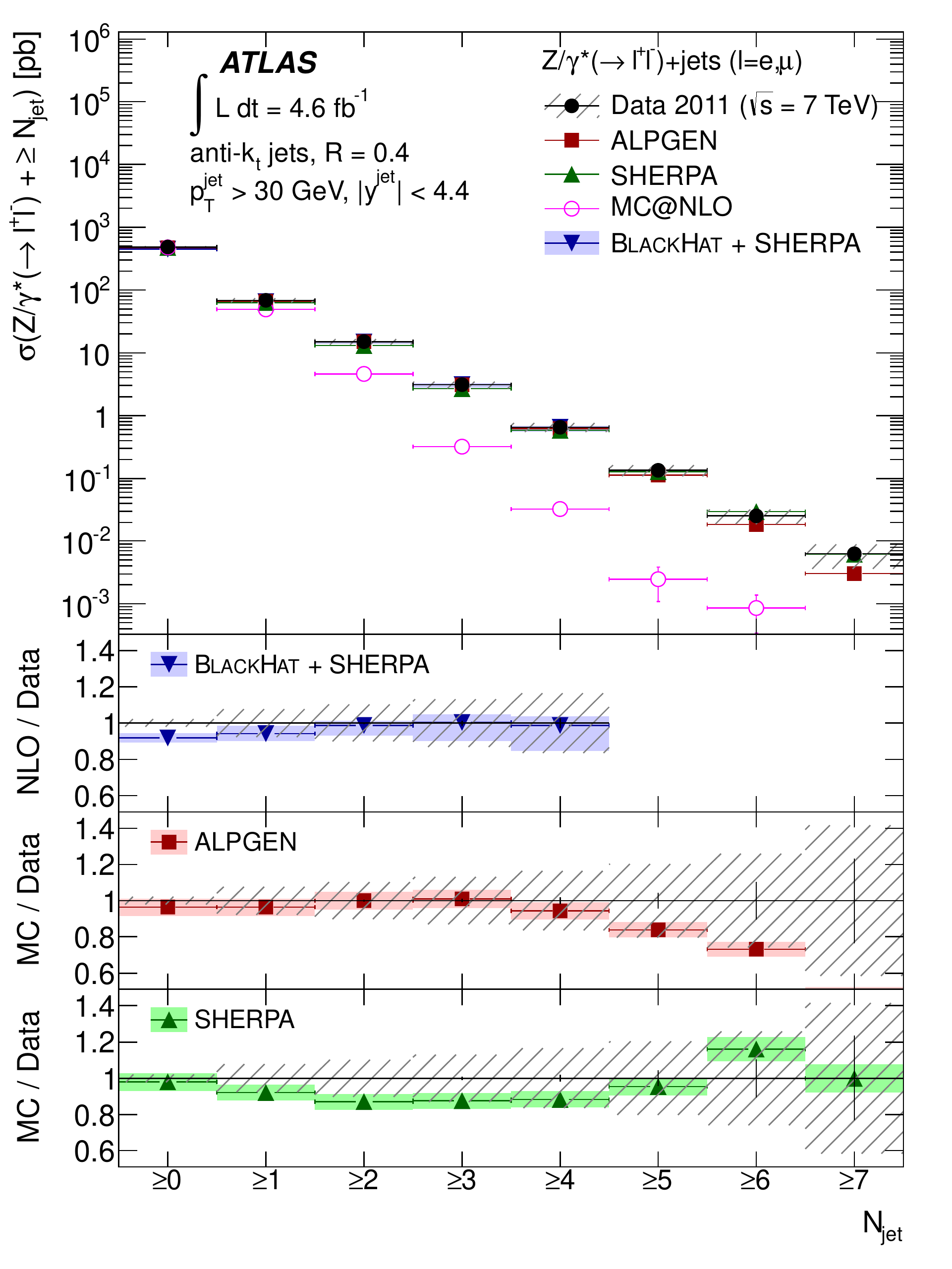}}
\caption{Measured cross section for $Z(\to\ell\ell)$+jets as a function of the inclusive jet multiplicity, $N_{\rm jet}$. From Ref.~\cite{zjets}. \label{fg:zjets}}
\end{figure}

Measurements of the production of jets of particles in association with a $Z$ boson in $pp$ collisions at
$\sqrt{s} = 7~\tev$ are presented in Ref.~\cite{zjets}, using data corresponding to an integrated luminosity of 4.6~\ifb\ collected by the ATLAS at the LHC. Inclusive and differential jet cross sections in $Z$ events, with $Z$ decaying into electron or muon pairs, are measured for jets with transverse momentum $\pt > 30~\gev$ and rapidity $|\eta| < 4.4$. The results are compared to next-to-leading-order perturbative QCD calculations, and to predictions from different Monte Carlo (MC) generators based on leading-order and next-to-leading-order matrix elements supplemented by parton showers.

Measured cross section for $Z(\to\ell\ell)$+jets as a function of the inclusive jet multiplicity, $N_{\rm jet}$ are depicted in Fig.~\ref{fg:zjets}. The data are compared to NLO pQCD predictions from BH+SHERPA corrected to the particle level, and the ALPGEN, SHRPA and MC@NLO event generators (see legend for details). The error bars indicate the statistical uncertainty on the data, and the hatched (shaded) bands the statistical and systematic uncertainties on data (prediction) added in quadrature. This analysis confirms the Poisson scaling for exclusive bins in the high-\pt\ regime.

\section{Top physics}\label{sc:top}

The study of the top quark is important for High Energy Physics for a number of reasons. It is the heaviest fermion ---near the electroweak (EW) symmetry-breaking scale---, therefore it features a large coupling to the Higgs boson. The top-quark production cross sections provide a test of QCD, since it is produced at very small distances. The top decays before hadronisation, allowing the study of spin characteristics (production mechanisms) and the $W$-helicity measurement (test of EW V-A structure). Its cross sections are sensitive to new physics, e.g.\ through the decay $t\to H^+b$. Besides these, it is an important background for Higgs studies and most searches for BSM scenarios.

The measurement of the top quark pair ($t\bar{t}$) inclusive production cross section in $pp$ collisions at $\sqrt{s} = 8~\tev$ is discussed in Ref.~\cite{ttbar}. The analysis has been done in the lepton-plus-jets final state in a dataset corresponding to an integrated luminosity of 5.8~\ifb. A multivariate technique and $b$-jet identification were employed to separate the signal $t\bar{t}$ events from the various backgrounds. The inclusive $t\bar{t}$ production cross section is measured to be $\sigma_{t\bar{t}} = 241 \pm 2 (\text{stat}) \pm 31 (\text{syst}) \pm 9 (\text{lumi})~{\rm pb}$ and is in good agreement with the theoretical prediction $\sigma_{t\bar{t},{\rm th}} = 238^{+22}_{-24}~{\rm pb}$. The overview of the $t\bar{t}$ cross section measurement for various centre-of-mass energies is presented in Fig.~\ref{fg:ttbar}.

\begin{figure}[htb]
{\includegraphics[width=0.7\textwidth]{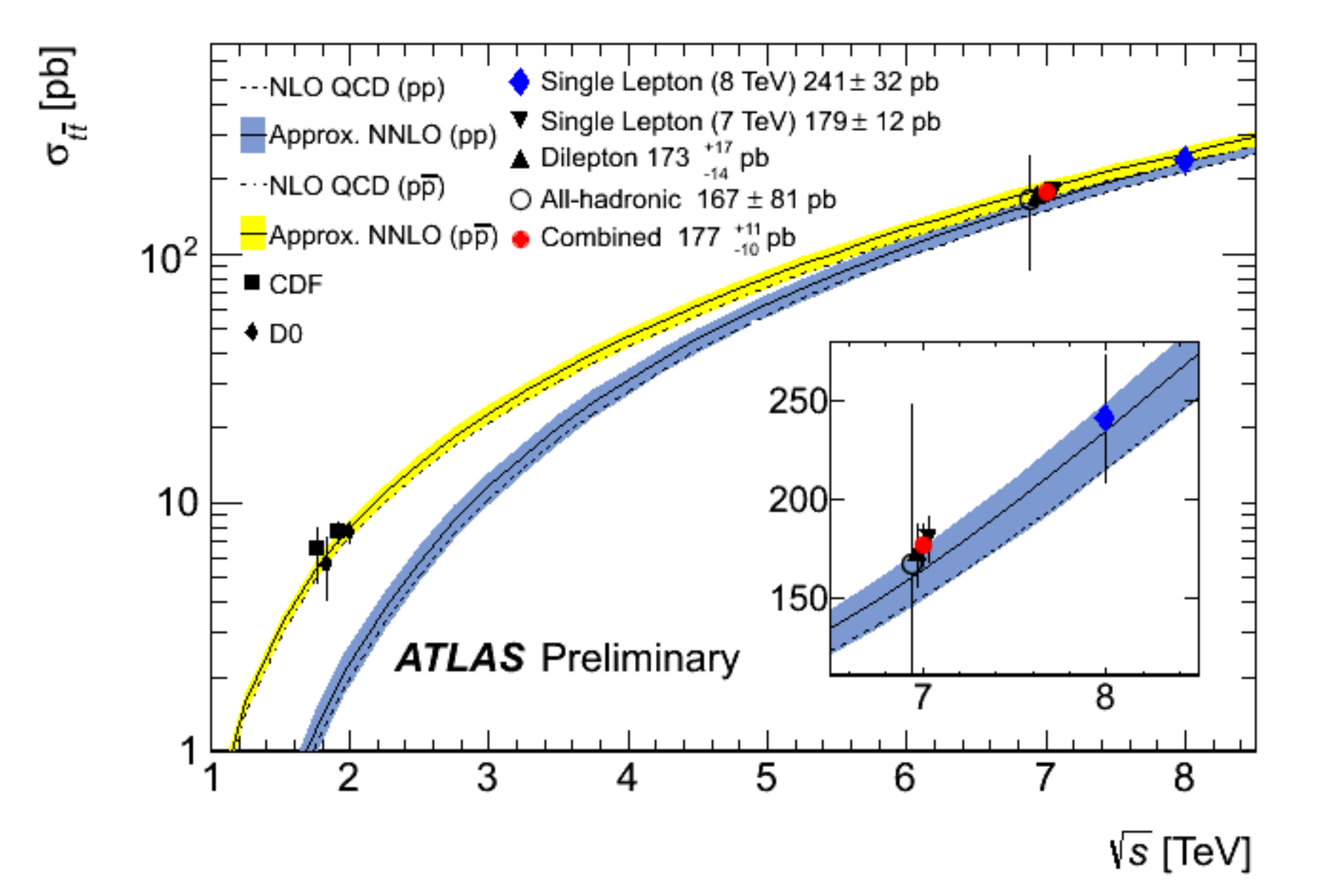}}
\caption{Summary plot showing the top pair production cross section as a function of the proton-(anti)proton centre-of-mass energy. The experimental results in the various top decay channels (and their combination) at 7 ~\tev\ and the recent result at 8~\tev\ are compared to an approximate NNLO QCD calculation based on Hathor 1.2. From Ref.~\cite{summary-results}. \label{fg:ttbar}}
\end{figure}

The top quark mass, on the other hand, has been measured using the template method in the channel $t\bar{t}\to\text{lepton}+\text{jets}$ ($\text{lepton}=e,\mu$) based on ATLAS data recorded in the year 2011~\cite{t-mass}. The data were taken at a proton-proton centre-of-mass energy of $\sqrt{s} = 7~\tev$ and correspond to an integrated luminosity of about 4.7~\ifb. This analysis uses a three-dimensional template technique which determines the top-quark mass together with a global jet energy scale factor (JSF), and a relative $b$-jet to light-jet energy scale factor (bJSF), where light jets refers to $u, d, c, s$ quark jets. The top quark mass is measured to be: $m_{\text{top}}= 172.31 \pm 0.23 (\text{stat}) \pm 0.27 (\text{JSF}) \pm 0.67 (\text{bJSF}) \pm 1.35 (\text{syst})~\gev$ or, equivalently, $m_{\text{top}}= 172.31 \pm 0.75 (\text{stat + JSF + bJSF}) \pm 1.35 (\text{syst})~\gev$, where the uncertainties labelled JSF and bJSF refer to the statistical uncertainties on $m_{\text{top}}$ induced by the \emph{in-situ} determination of these scale factors. The summary of all ATLAS direct $m_{\text{top}}$ measurements compared with the Tevatron and CMS ones is shown in Fig.~\ref{fg:t-mass}. 

\begin{figure}[htb]
{\includegraphics[width=0.9\textwidth]{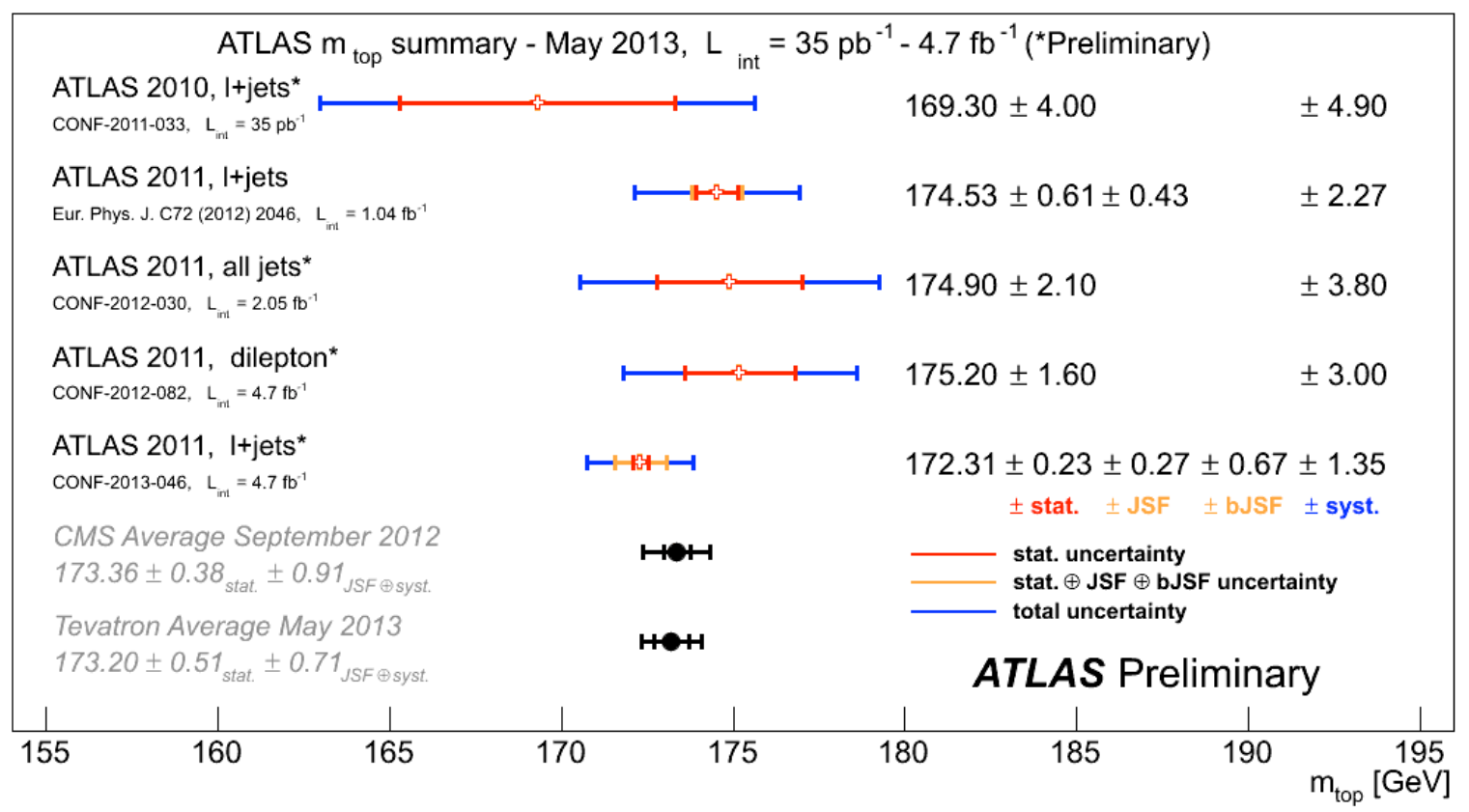}}
\caption{Summary of the ATLAS direct $m_{\text{top}}$ measurements. The results are compared to the 2013 Tevatron and 2012 CMS $m_{\text{top}}$ combinations. For each measurement, the statistical uncertainty, the JSF and bJSF contributions (when applicable) as well as the sum of the remaining uncertainties are reported separately. From Ref.~\cite{summary-results}. \label{fg:t-mass}}
\end{figure}

The top quark polarisation in $t\bar{t}$ events using the lepton-plus-jets final state, where one $W$ boson decays leptonically and the other hadronically, has been measured with ATLAS~\cite{ttbar-pol}. The decay of the $t\bar{t}$ pair is fully reconstructed using a likelihood method in order to calculate the rest frame of the leptonically decaying top quark. A template fit to the distribution of lepton polar angles in the parent top quark's rest frame is used to extract the fraction of positively polarised top quarks. The full 2011 ATLAS 7~\tev\ centre of mass energy $pp$ collisions dataset from the LHC (4.66~\ifb) is used to perform the measurement. The fraction of positively polarised top quarks is found to be $f=0.470\pm0.009(\text{stat})\,^{+0.023}_{-0.032}(\text{syst})$, compatible with the Standard Model prediction of $f=0.5$.

Recent measurements of the $W$-boson polarisation in top-quark decays performed by the ATLAS and CMS Collaborations have been combined in Ref.~\cite{w-pol}. The measurements are based on proton-proton collision data corresponding to integrated luminosities ranging from 35~\ipb\ to 2.2~\ifb\ produced at the LHC at a centre-of-mass energy of $\sqrt{s} = 7~\tev$. The results are quoted as helicity fractions, i.e.\ the fractions of events which contain $W$ bosons with longitudinal and left-handed polarisation. The combined helicity fractions are
\begin{eqnarray}
F_0 &=& 0.626 \pm 0.034 (\text{stat}) \pm 0.048 (\text{syst}), \\
F_L &=& 0.359 \pm 0.021 (\text{stat}) \pm 0.028 (\text{syst}),
\end{eqnarray}
which are in agreement with predictions from NNLO QCD. The fraction of $W$ bosons with right-handed polarisation is calculated assuming the sum of all fractions to be unity:
\begin{equation}
F_R = 0.015 \pm 0.034,
\end{equation}
where the uncertainty includes the statistical and systematic uncertainties. Exclusion limits on anomalous $Wtb$ couplings are derived from these results.

\section{$B$ physics}\label{sc:bphys}

Weak decays of hadrons containing heavy quarks are employed for tests of the Standard Model and measurements of its parameters. In particular, they offer the most direct way to determine the weak mixing angles, to test the unitarity of the Cabibbo-Kobayashi-Maskawa (CKM) matrix, and to explore the physics of $CP$ violation. The latter may also provide some hints about New Physics beyond the Standard Model. On the other hand, hadronic weak decays also serve as a probe of that part of strong-interaction phenomenology which is least understood: the confinement of quarks and gluons inside hadrons. In ATLAS beauty-physics analyses also include heavy quarkonia and $D$-hadron production in a variety of channels and analyses. 

A limit on the branching fraction of $B_s^0 \to\mu\mu$ has been set using an integrated luminosity of 4.9~\ifb\ collected in 2011 by the ATLAS detector~\cite{b-mumu}. The decay $B^{\pm}\to J/\psi K^{\pm}$, with $J/\psi\to\mu^{+}\mu^{-}$, was used as a reference channel for the normalisation of the integrated luminosity, acceptance and efficiency. The final selection was based on a multivariate analysis which is trained on MC, leaving the events in the sidebands to be used for optimisation and background estimation. Furthermore, this result profits from an improved event reconstruction and the use of multi-dimensional unbinned maximum likelihood fits for the extraction of the reference channel yield, with respect to previous results. An upper limit on $BR(B_s^0 \to\mu\mu) < 1.5~(1.2) \times10^{-8}$ at 95\% (90\%) confidence level (CL) has been set, as observed in Fig.~\ref{fg:b-mumu}.

\begin{figure}[htb]
{\includegraphics[width=0.7\textwidth]{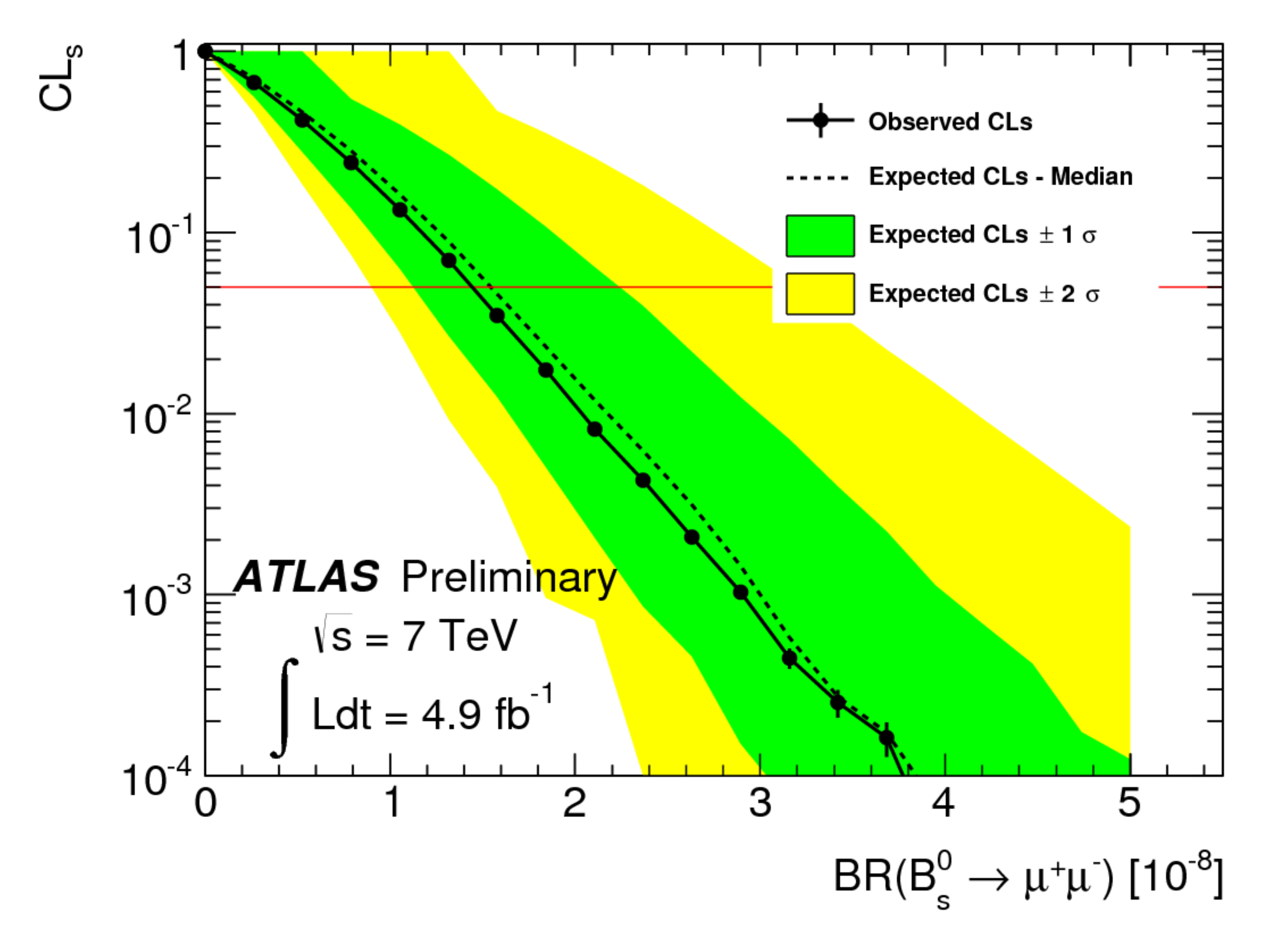}}
\caption{Observed CLs (circles) as a function of $B_s^0 \to\mu\mu$. The 95\% CL limit is indicated by the horizontal (red) line. The dark (green) and light (yellow) bands correspond to the $\pm1\sigma$ and $\pm2\sigma$ ranges of the background-only pseudo-experiments with the median of the expected CLs given by the dashed line. From Ref.~\cite{b-mumu}. \label{fg:b-mumu}}
\end{figure}

Using 4.9~\ifb\ of integrated luminosity taken at $\sqrt{s} = 7~\tev$ by the ATLAS experiment, $B_d^0 \to K^{*0}\mu^+\mu^-$ events have been reconstructed and the angular distribution of their final state particles hase been measured~\cite{b-angular}. The forward-backward asymmetry $A_{FB}$ and the $K^{*0}$ longitudinal polarisation $F_L$ have been measured as a function of the dimuon invariant mass squared, $q^2$. The results obtained on $A_{FB}$ and $F_L$ are mostly consistent with theoretical predictions and measurements performed by other experiments. The results for $F_L$ in the low-$q^2$ bins slightly deviate from Standard Model expectations, as reflected in Fig.~\ref{fg:b-angular}.

\begin{figure}[htb]
{\includegraphics[width=0.7\textwidth]{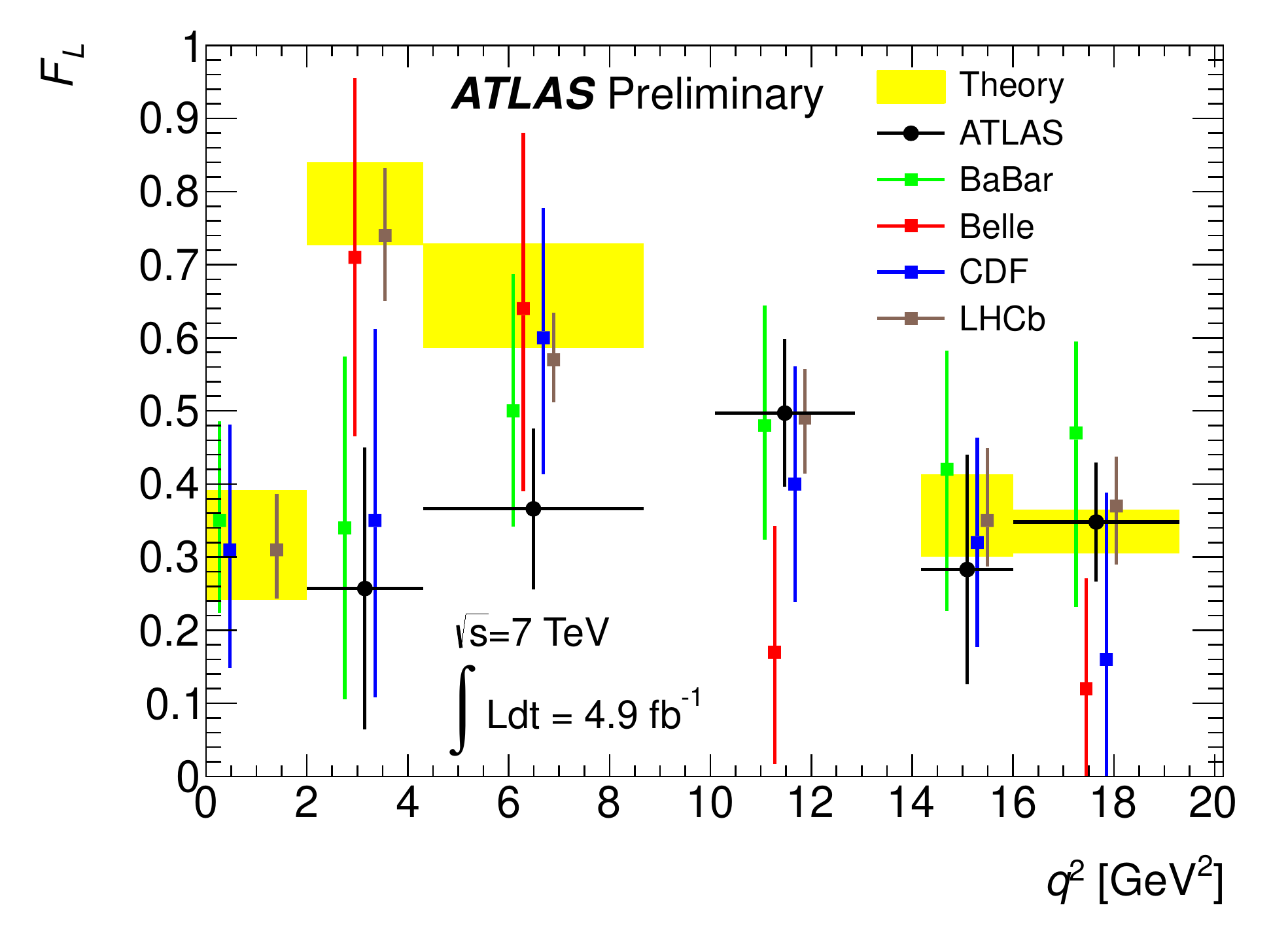}}
\caption{Fraction of longitudinal polarised $K^{*0}$ mesons $F_L$ as a function of $q^2$ measured by ATLAS (black dots). In each $q^2$ bin, ordered from right to left, results of other experiments are shown as coloured squares: BaBar, Belle, CDF and LHCb. All errors including statistical and systematic uncertainties. The experimental results are compared to theoretical Standard Model predictions including theoretical uncertainties. From Ref.~\cite{b-angular}. \label{fg:b-angular}}
\end{figure}

\section{Heavy-ion runs and results}\label{sc:hi}

During the operation in the years 2009-2013, besides collisions of proton beams, the LHC also delivered Pb+Pb collisions and $p$+Pb collisions, at the centre-of-mass energy $\sqrt{s_{NN}} = 2.76~\tev$ and $\sqrt{s_{NN}} = 5.02~\tev$, respectively. Studies of heavy-ion collisions explore complicated systems created in the volumes much larger than the size of the proton under the conditions of extreme energy density. Such conditions are similar to those at an early phase of the Universe evolution, when the matter was in the state of strongly interacting quark-gluon plasma. Results of nuclei collisions depend on the size of the volume and its shape, which are directly connected with the smallest distance of the nuclei centres during the collision. Commonly, centrality of the nuclei collisions is denoted by the percentage of the cross section (starting from the most central events) or is characterised by the number of nucleons participating in inelastic interactions or by the number of binary nucleon-nucleon collisions.

\begin{figure}[htb]
{\includegraphics[width=\textwidth]{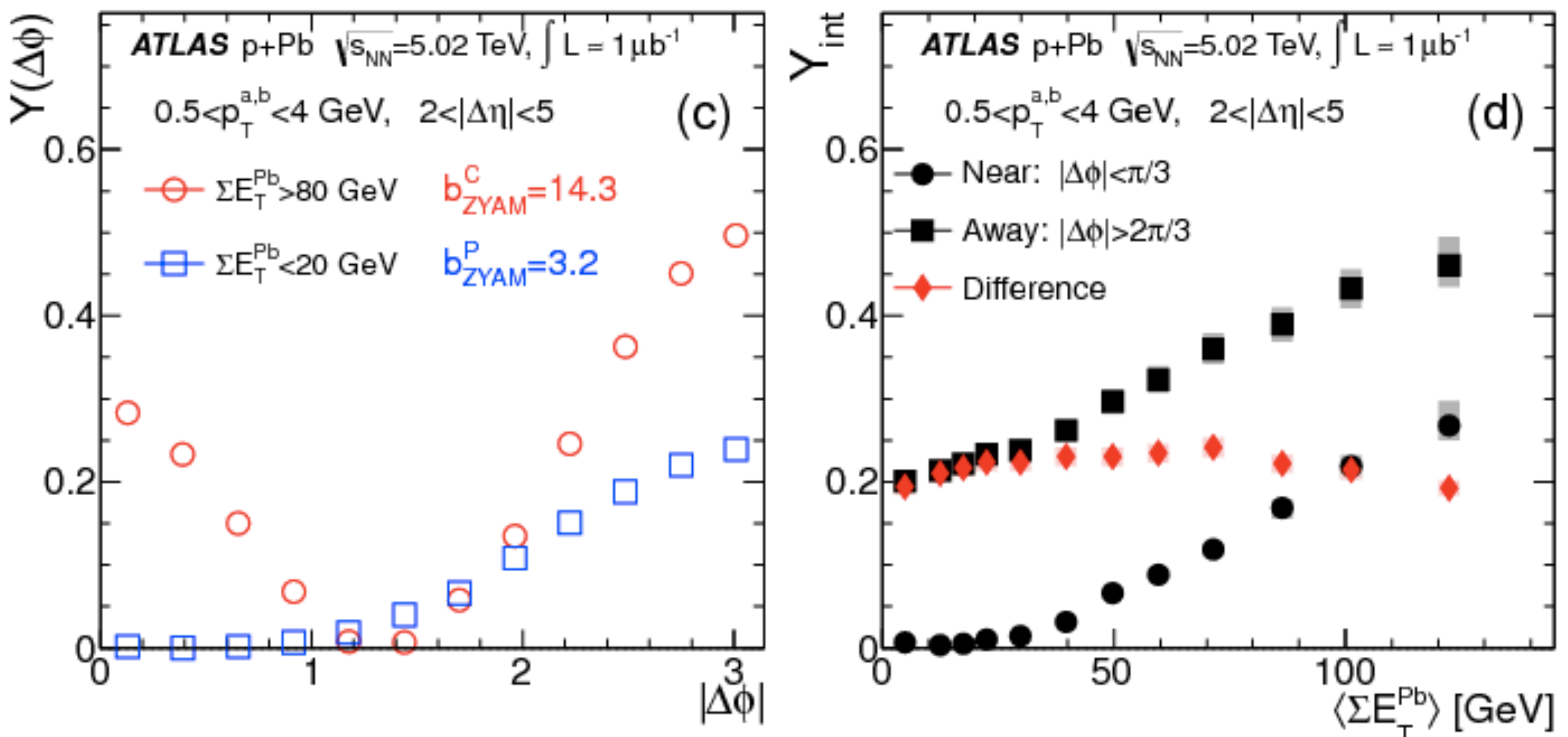}}
\caption{{\it Left:} Per-trigger yield $\Delta\phi$ distribution together with pedestal levels for peripheral $(b{\rm ^P_{ZYAM}})$ and central $(b{\rm ^C_{ZYAM}})$ events. {\it Right:} Integrated per-trigger yield as a function of $\sum E_{\rm T}^{\rm Pb}$ for pairs in $2<|\Delta\eta|<5$. The shaded boxes represent the systematic uncertainties, and the statistical uncertainties are smaller than the symbols. From Ref.~\cite{hi}. \label{fg:hi}}
\end{figure}

Two-particle correlations in relative azimuthal angle $\Delta\phi$ and pseudorapidity $\Delta\eta$ were measured in $\sqrt{s_{NN}} = 5.02~\tev$ $p$+Pb collisions using the ATLAS detector at the LHC~\cite{hi}. The measurements were performed using approximately 1~\imb\ of data as a function of \pt\ and the transverse energy $\sum E_{\rm T}^{\rm Pb}$ summed over $3.1 < \eta < 4.9$ in the direction of the Pb beam. The correlation function, constructed from charged particles, exhibits a long-range $(2<|\Delta\eta|<5)$ near-side $(\Delta\phi\sim 0)$ correlation that grows rapidly with increasing $\sum E_{\rm T}^{\rm Pb}$. A long-range away-side $(\Delta\phi\sim\pi)$ correlation, obtained by subtracting the expected contributions from recoiling dijets and other sources estimated using events with small $\sum E_{\rm T}^{\rm Pb}$, is found to match the near-side correlation in magnitude, shape (in $\Delta\eta$ and $\Delta\phi$) and $\sum E_{\rm T}^{\rm Pb}$ dependence, as shown in Fig.~\ref{fg:hi} (right). The resultant $\Delta\phi$ correlation is approximately symmetric about $\pi/2$, and is consistent with a $\cos(2\Delta\phi$) modulation for all $\sum E_{\rm T}^{\rm Pb}$ ranges and particle \pt, as shown in Fig.~\ref{fg:hi} (left). The amplitude of this modulation is comparable in magnitude and \pt\ dependence to similar modulations observed in heavy-ion collisions, suggestive of final-state collective effects in high multiplicity events. 

\section{Higgs boson discovery and properties}\label{sc:higgs}

The Brout-Englert-Higgs mechanism~\cite{beh} plays a central role in the unification of the electromagnetic and weak interactions by providing mass to the $W$ and $Z$ intermediate vector bosons without violating local gauge invariance. Within the Standard Model, the Higgs mechanism is invoked to break the electroweak symmetry; it implies the existence of a single neutral scalar particle, the Higgs boson. In July 2012 the ATLAS~\cite{higgs-disc1} and CMS~\cite{higgs-disc2} Collaborations announced the discovery of a new particle, with evidence of decays to photons as well as $Z$ and $W$ bosons. In the following year the LHC has produced more data, allowing for a more detailed understanding of this new resonance. With this dataset, the ATLAS experiment has studied the properties of the particle, including its spin, mass, and couplings to SM particles, and found these properties to be consistent with those of the Standard Model Higgs boson.
 
Crucial experimental aspects of ATLAS for the observation of the diphoton decay of the Higgs boson are the excellent $\gamma\gamma$ mass resolution to observe the narrow signal peak above irreducible background and  the powerful $\gamma/\text{jet}$ separation to suppress  $\gamma$-jet and jet-jet background with $\text{jet}\to\pi^0$ faking single photons. Measurements of the properties~\cite{h-gg-prop}, the spin~\cite{h-gg-spin} and the cross section~\cite{h-gg-xsec} of the Higgs boson in the $H\to\gamma\gamma$ channel have been performed using the complete dataset of $\sim20~\ifb$ of $pp$ collision data at a centre-of-mass energy of 8~\tev. 

The diphoton invariant-mass distributions for the $N_{\rm jets}\geq3$ bin is shown in the left panel of Fig.~\ref{fg:h-gg-4l}~\cite{h-gg-xsec}. The curves show the results of the single simultaneous fit to data for all $N_{\rm jets}$ bins. The red line is the combined signal and background, and the dashed line shows the background. The difference of the two curves is the extracted signal yield. The bottom inset displays the residuals of the data with respect to the fitted background component, and the dotted red line corresponds to the signal. The lower local probability of the background fluctuating beyond the observation in the data at a particular $m_H$ ($p_0$ value) of $\sim10^{-13}$ ($7.4\sigma$ significance) was found at $m_H =126.5~\gev$. The best-fit $m_H$ value was found to be $126.8\pm0.2\text{(stat)}\pm0.7\text{(syst)}~\gev$. The dominant contribution to the systematic uncertainty on $m_H$ comes from the uncertainties on the photon energy scale. At the best-fit value of $m_H =126.8~\gev$, the signal strength, i.e.\ the ratio of the observed cross section to the expected SM cross section, was found to be $\mu=1.65_{-0.30}^{+0.34}$.

\begin{figure}[htb]
{\includegraphics[width=0.55\textwidth]{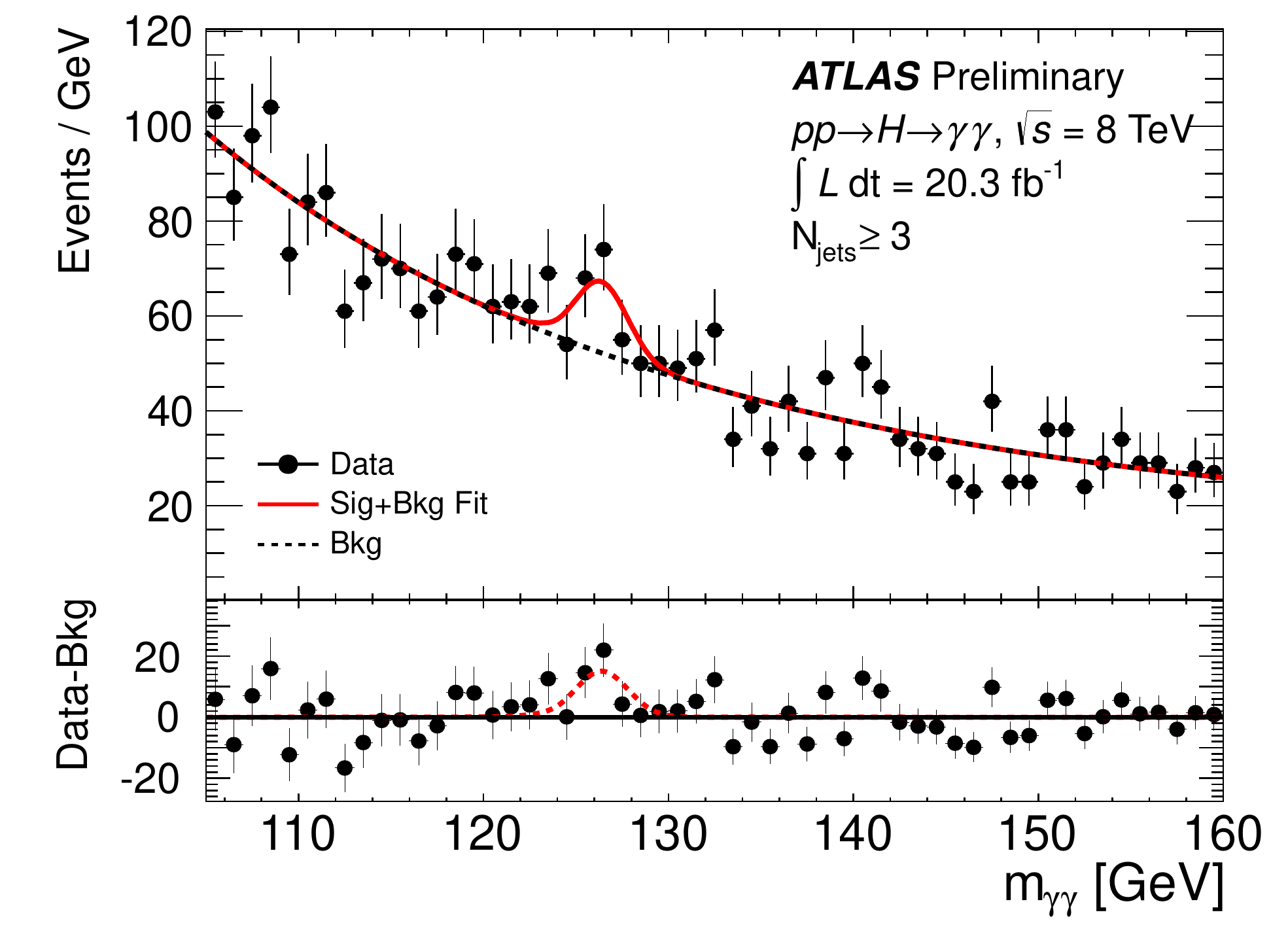}}
{\includegraphics[width=0.45\textwidth]{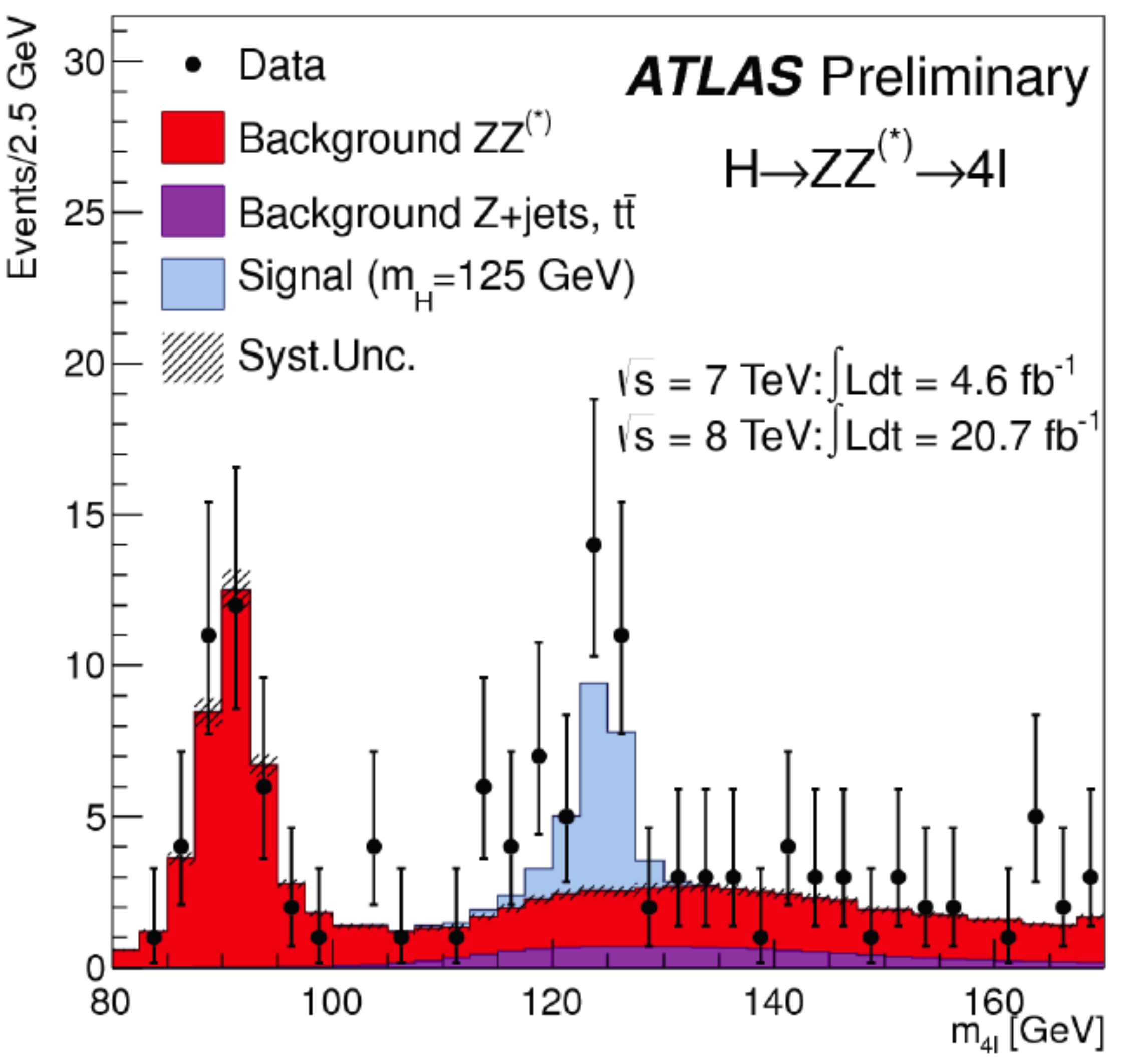}}
\caption{{\it Left:} Diphoton invariant-mass distribution for the $N_{\rm jets}\geq3$ bin. From Ref.~\cite{h-gg-xsec}. {\it Right:} The distributions of the four-lepton invariant mass, $m_{4\ell}$, compared to the background expectation for the combined $\sqrt{s}=8~\tev$ and $\sqrt{s}=7~\tev$ data sets in the mass range $80-170~\gev$. From Ref.~\cite{h-4l}. \label{fg:h-gg-4l}}
\end{figure}

Also known as the ``golden'' decay channel, $H\to ZZ^{(*)}\to 4\ell$ features by far the cleanest sample $(S/N\sim 1)$ due to its unique signature of two high energy lepton pairs. However for $m_H$ below 180~\gev, the $ZZ$ decay is sub-threshold, thus forcing one of the $Z^0$ off-shell, thereby significantly reducing the branching ratio. The ATLAS detector was conceived to measure high-energy muons and electrons with great precision, also incorporating recognition of fast leptons already at the first (hardware) stage of the trigger. The distributions of the four-lepton invariant mass, $m_{4\ell}$, for the selected candidates compared to the background expectation in the mass range $80-170~\gev$ are shown in Fig.~\ref{fg:h-gg-4l}~\cite{h-4l}. The signal expectation for the $m_H=125~\gev$ hypothesis is also shown. The resolution of the reconstructed Higgs boson mass is dominated by detector resolution at low $m_H$ values and by the Higgs boson width at high $m_H$. The results are based on 4.6~\ifb\ of 7~\tev\ $pp$ collision data combined with 20.7~\ifb\ of 8~\tev. A clear excess of events over the background is observed at $m_H = 124.3~\gev$ in the combined analysis of the two datasets with a significance of $6.6\sigma$, corresponding to a background fluctuation probability of $2.7\times10^{-11}$. The mass of the Higgs-like boson is measured to be $m_H = 124.3 _{-0.5}^{+0.6}\text{(stat)} _{-0.3}^{+0.5}\text{(syst)}~\gev$ in this channel, and the signal strength at this mass is found to be $\mu = 1.7_{-0.4}^{+0.5}$.

The full datasets recorded by ATLAS in 2011 and 2012, corresponding to an integrated luminosity of up to 25~\ifb\ at $\sqrt{s} = 7~\tev$ and $\sqrt{s} = 8~\tev$, have been analysed to determine several properties of the recently discovered Higgs boson using the $H\to\gamma\gamma$, $H\to ZZ^*\to4\ell$ and $H\to WW^*\to\ell\nu\ell\nu$ decay modes~\cite{h-couplings}. The reported results include measurements of the mass and signal strength, evidence for production through vector-boson fusion, and constraints on couplings to bosons and fermions as well as on anomalous contributions to loop-induced processes. 

\begin{figure}[htb]
{\includegraphics[width=0.63\textwidth]{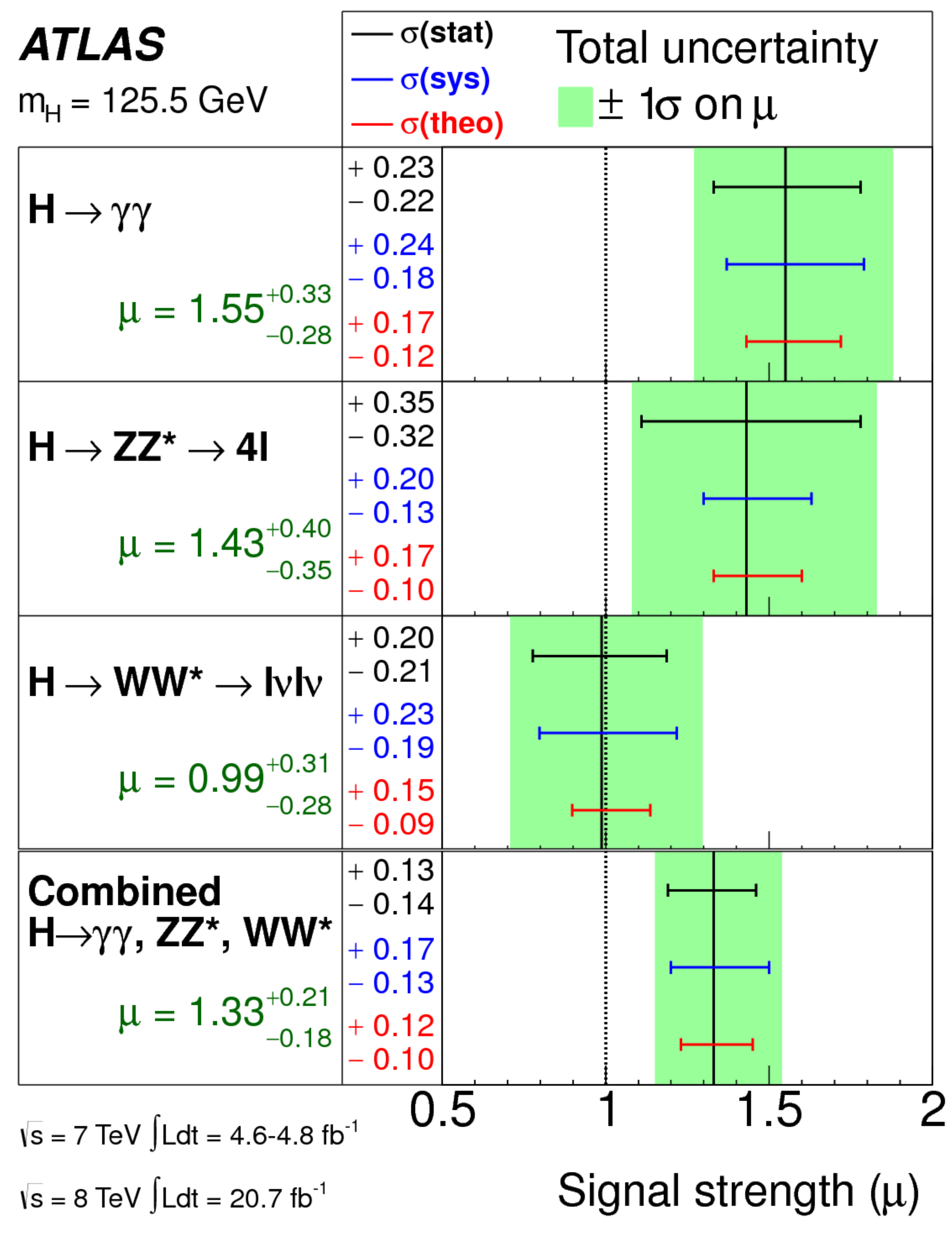}}
\caption{Measured production strengths for a Higgs boson of mass $m_H =125.5~\gev$, normalised to the SM expectations, for diboson final states and their combination. From Ref.~\cite{h-couplings}. \label{fg:h-signal-mu}}
\end{figure}

The measured production strengths for a Higgs boson of mass $m_H =125.5~\gev$, normalised to the SM expectations, for diboson final states and their combination is visible in Fig.~\ref{fg:h-signal-mu}. The best-fit values are indicated by the solid vertical lines. The total $\pm1\sigma$ uncertainty is indicated by the shaded band, with the individual contributions from the statistical uncertainty (top), the total (experimental and theoretical) systematic uncertainty (middle), and the theoretical uncertainty (bottom) on the signal cross section (from QCD scale, PDF, and branching ratios) shown as superimposed error bars. The overall compatibility between the signal strengths measured in the three final states and the SM predictions is about 14\%, with the largest deviation $(\sim1.9\sigma)$ observed in the $H\to\gamma\gamma$ channel~\cite{h-couplings}.

Good consistency between the measured and expected signal strengths is also found for the various categories of the $H\to\gamma\gamma$, $H\to ZZ^*\to4\ell$ and $H\to WW^*\to\ell\nu\ell\nu$ analyses, which are the primary experimental inputs to the fit. If the preliminary $H\to\tau\tau$~\cite{h-tautau} and $H\to b\bar{b}$ ~\cite{h-bb} results, for which only part of the 8~\tev\ dataset is used (13~\ifb), were included, the combined signal strength would be $\mu = 1.23 \pm 0.18$~\cite{h-couplings}.

\begin{figure}[htb]
{\includegraphics[width=0.65\textwidth]{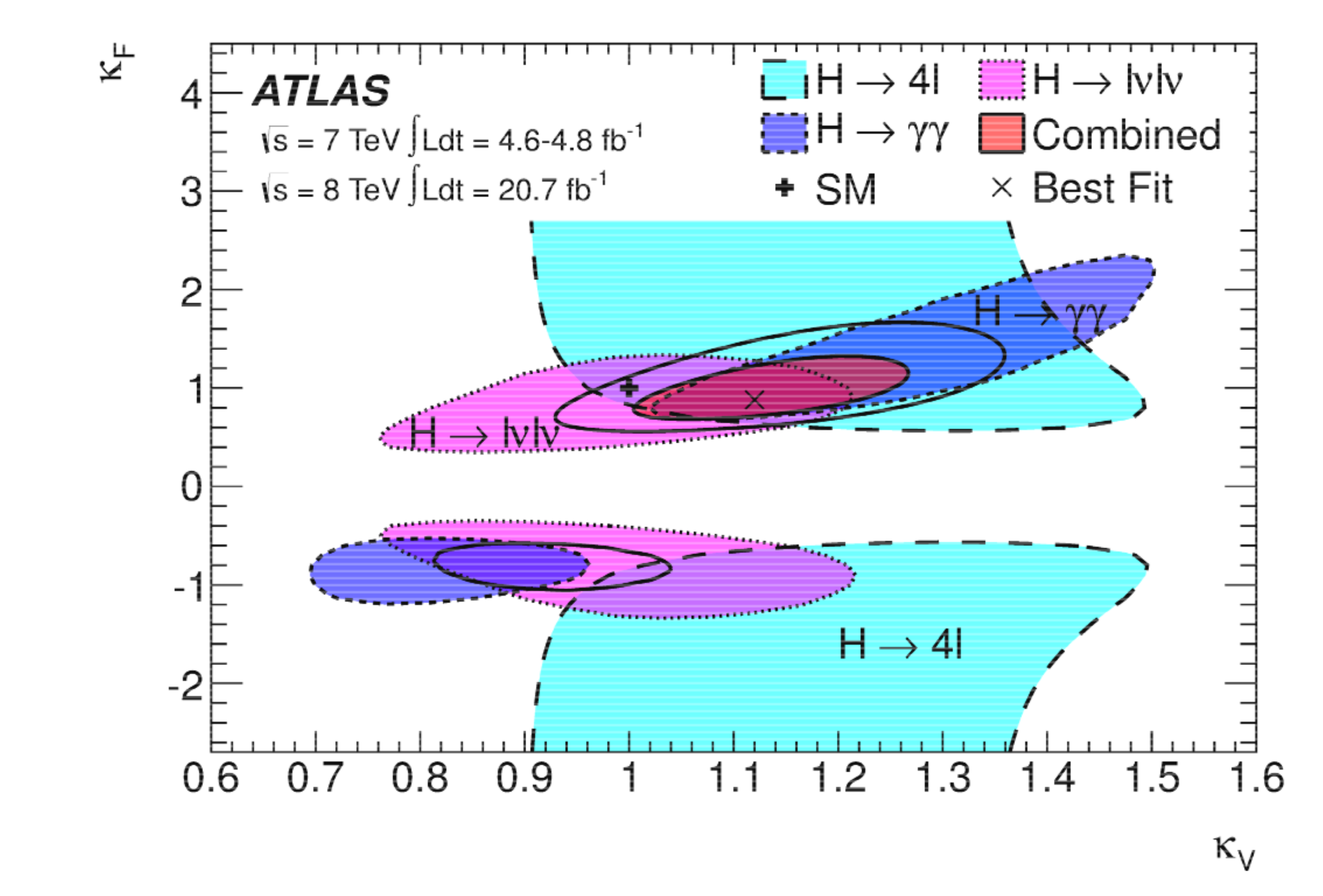}}
\caption{Likelihood contours (68\% CL) of the coupling scale factors $\kappa_F$ and $\kappa_V$ for fermions and bosons, as obtained from fits to the three individual channels and their combination (for the latter, the 95\% CL contour is also shown). The best-fit result ($\times$) and the SM expectation ($+$) are also indicated. From Ref.~\cite{h-couplings}. \label{fg:h-couplings}}
\end{figure}

The benchmark model for the Higgs couplings presented here assumes one coupling scale factor for fermions, $\kappa_F$, and one for bosons, $\kappa_V$; in this scenario, the $H\to\gamma\gamma$ and $gg\to H$ loops and the total Higgs boson width depend only on $\kappa_F$ and $\kappa_V$, with no contributions from physics beyond the Standard Model. The strongest constraint on $\kappa_F$ comes indirectly from the $gg\to H$ production loop. Figure~\ref{fg:h-couplings} shows the results of the fit to the data for the three channels and their combination. Since only the relative sign of $\kappa_F$ and $\kappa_V$ is physical, in the following $\kappa_V > 0$ is assumed. Some sensitivity to this relative sign is provided by the negative interference between the $W$-boson loop and $t$-quark loop in the $H\to\gamma\gamma$ decay. The data prefer the minimum with positive relative sign, which is consistent with the SM prediction, but the local minimum with negative sign is also compatible with the observation (at the $\sim2\sigma$ level). The two-dimensional compatibility of the SM prediction with the best-fit value is 12\%.

\begin{figure}[htb]
{\includegraphics[width=0.45\textwidth]{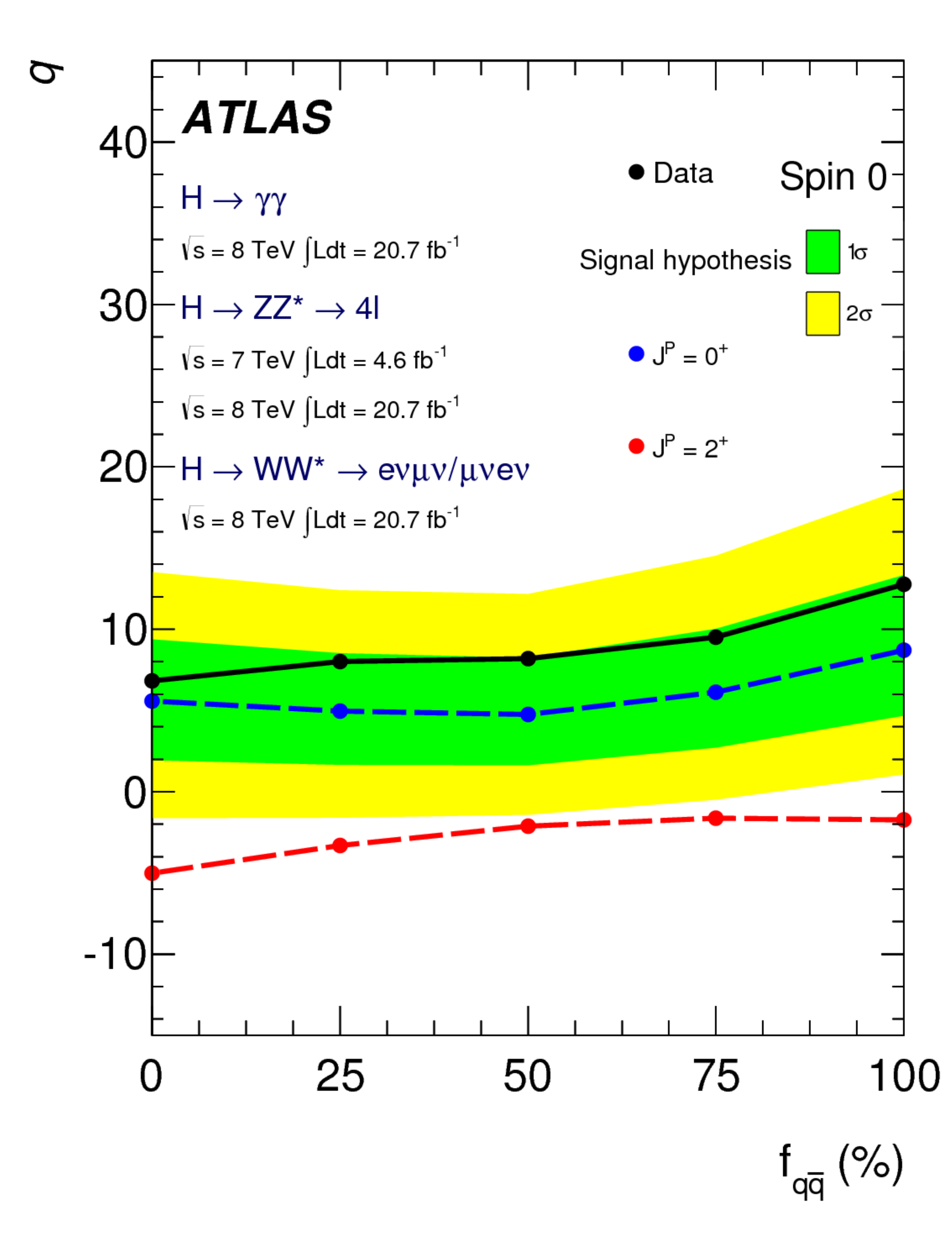}}
{\includegraphics[width=0.45\textwidth]{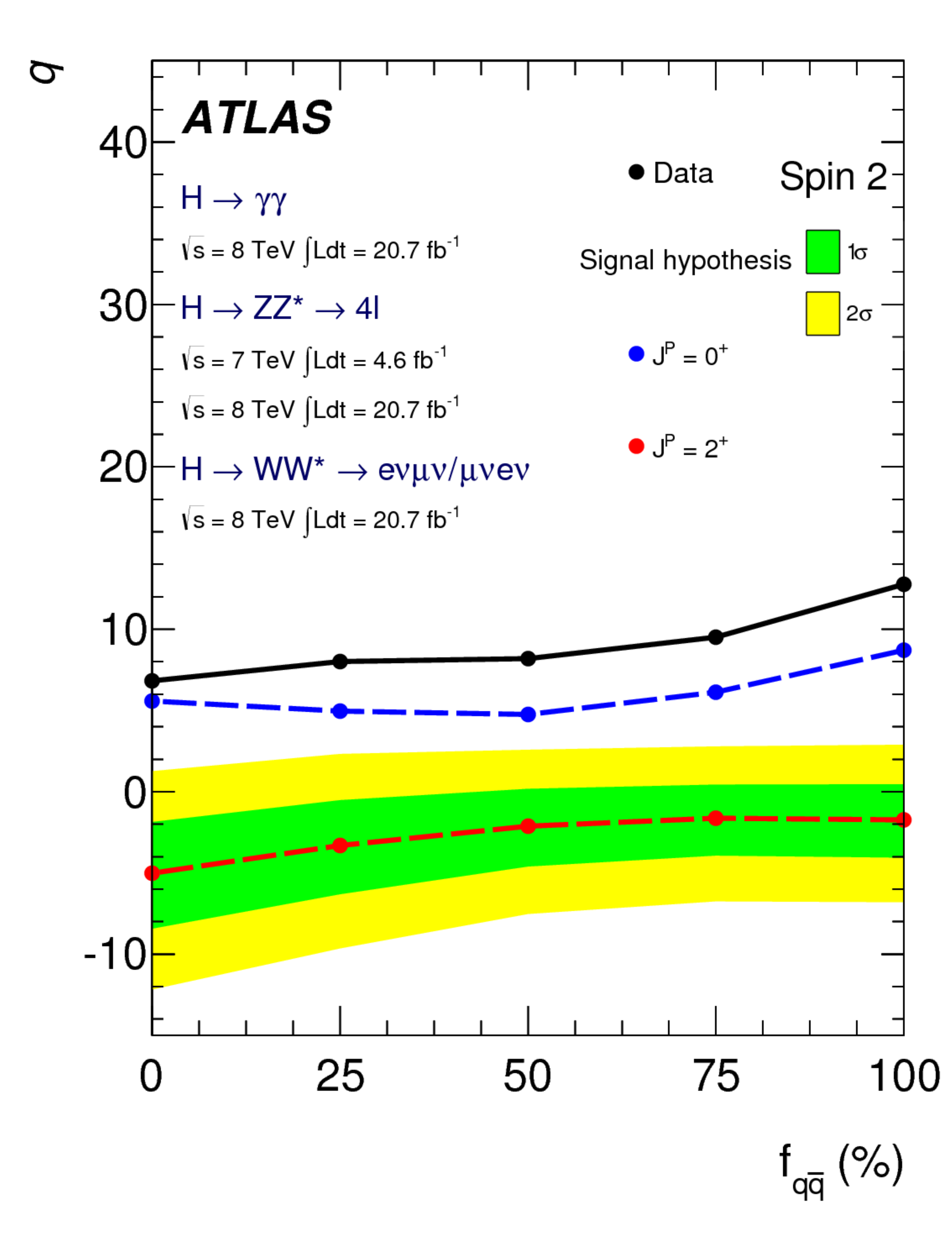}}
\caption{Expected distributions of the log-likelihood ratio $q$, for the combination of channels as a function of the fraction of the $q\bar{q}$ spin-2 production mechanism. The green and yellow bands represent, respectively, the $1\sigma$ and $2\sigma$ bands for $0^+$ (left) and for $2^+$ (right). From Ref.~\cite{h-spin}. \label{fg:h-spin}}
\end{figure}

The Standard Model $J^P=0^+$ hypothesis for the Higgs boson has been compared to alternative spin/parity hypotheses using 8~\tev\ (20.7~\ifb) and 7~\tev\ (4.6~\ifb) proton-proton collision data collected by ATLAS~\cite{h-spin}. The Higgs boson decays $H\to\gamma\gamma$, $H\to ZZ^*\to4\ell$ and $H\to WW^*\to\ell\nu\ell\nu$ have been used to test several specific alternative models, including $J^P = 0^-, 1^+, 1^-$ and a graviton-inspired $J^P = 2^+$ model with minimal couplings to SM particles. The data favour the Standard Model quantum numbers of $J^P = 0^+$. The $0^-$ hypothesis is rejected at 97.8\% CL by using the $H\to ZZ^*\to4\ell$ decay alone. The $1^+$ and $1^-$ hypotheses are rejected with a CL of at least 99.7\% by combining the $H\to ZZ^*\to4\ell$ and $H\to WW^*\to\ell\nu\ell\nu$ channels. Finally, the $J^P = 2^+$ model is rejected at more than 99.9\% CL by combining all three bosonic channels, independent of the assumed admixture of gluon-fusion and quark-antiquark production, as observed in Fig.~\ref{fg:h-spin}. All these alternative models are excluded without assumptions on the strength of the couplings of the Higgs boson to SM particles. These studies provide evidence for the spin-0 nature of the Higgs boson, with positive parity being strongly preferred.

\section{Hunt for supersymmetry}\label{sc:susy}

Supersymmetry (SUSY)~\cite{susy,lisboa} is an extension of the Standard Model which assigns to each SM field a superpartner field with a spin differing by a half unit. SUSY provides elegant solutions to several open issues in the SM, such as the hierarchy problem, the identity of dark matter, and grand unification.

SUSY searches in collider experiments typically focus on events with high transverse missing energy (\met) which can arise from (weakly interacting) Lightest Supersymmetric Particles (LSPs), in the case of $R$-parity conserving SUSY, or from neutrinos produced in LSP decays, when $R$-parity is broken. Hence, the event selection criteria of inclusive channels are based on large \met, no or few leptons ($e$, $\mu$), many jets and/or $b$-jets, $\tau$-leptons and photons. 

\begin{figure}[htb]
{\includegraphics[width=0.65\textwidth]{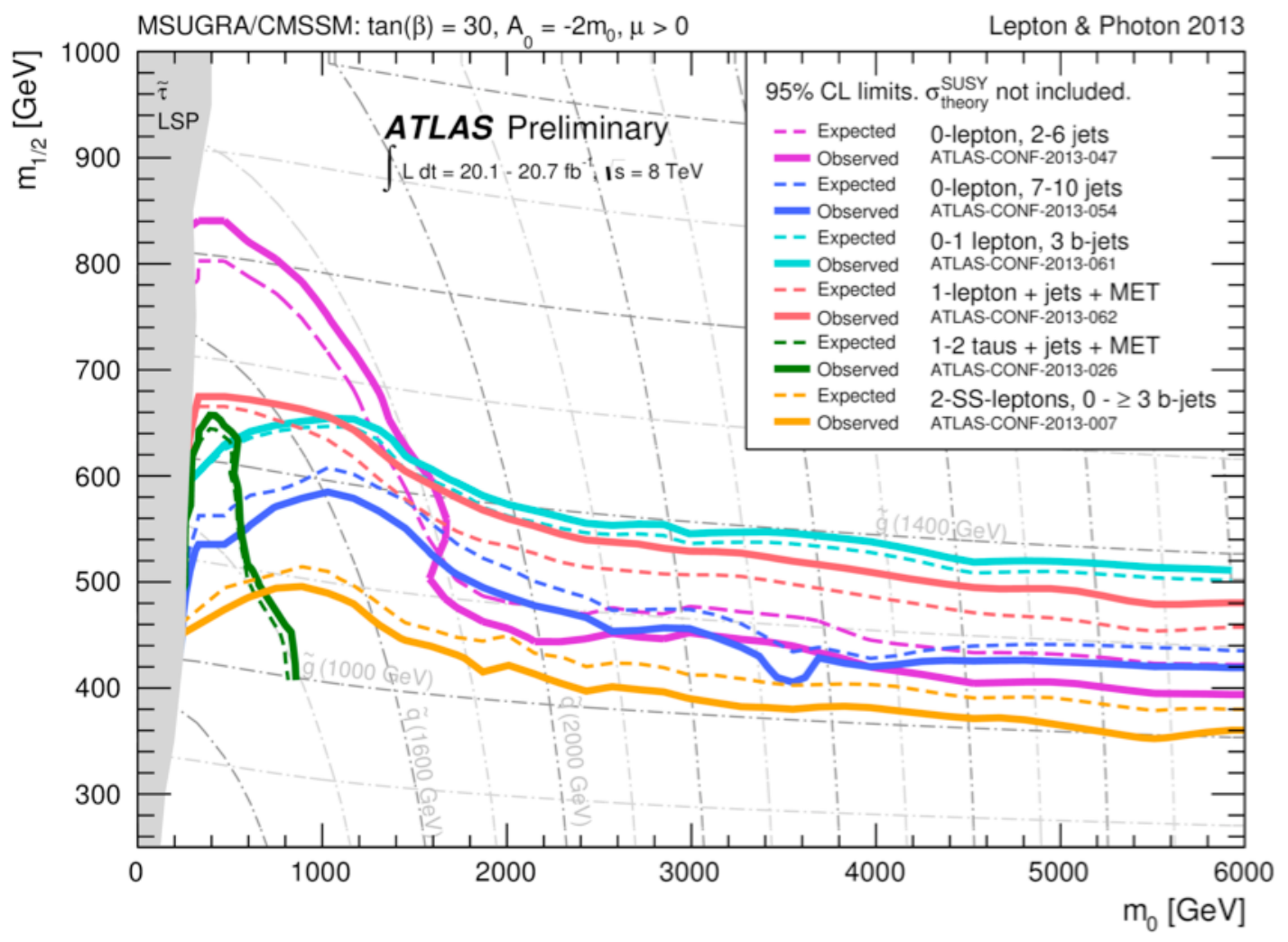}}
\caption{Exclusion limits at 95\% CL for 8~\tev\ analyses in the $(m_0,\,m_{1/2})$ plane for the MSUGRA/CMSSM model. From Ref.~\cite{summary-results}. \label{fg:msugra}}
\end{figure}

In the absence of deviations from SM predictions, limits for squark and gluino production are set. Figure~\ref{fg:msugra} illustrates the 95\% CL limits set by ATLAS under the minimal Supergravity (mSUGRA) model in the $(m_0,\,m_{1/2})$ plane. The remaining parameters are set to $\tan\beta = 30$, \mbox{$A_0 = -2\,m_0$}, $\mu > 0$, so as to acquire parameter-space points where the predicted mass of the lightest Higgs boson, $h_0$, is near $125~\gev$, i.e.\ compatible with the recently observed Higgs-like boson~\cite{higgs-disc1,higgs-disc2}. Exclusion limits are obtained by using the signal region with the best expected sensitivity at each point. 

The mixing of left- and right-handed gauge states which provides the mass eigenstates of the scalar quarks and leptons can lead to relatively light third-generation particles. Stop ($\tilde{t}_1$) and sbottom ($\tilde{b}_1$) with a sub-TeV mass are favoured by the naturalness argument, while the stau ($\tilde{\tau}_1$) is the lightest slepton in many models. Therefore these could be abundantly produced either directly or through  gluino production and decay. Such events are characterised by several energetic jets (some of them $b$-jets), possibly accompanied by light leptons, as well as high \met.

\begin{figure}[htb]
{\includegraphics[width=0.9\textwidth]{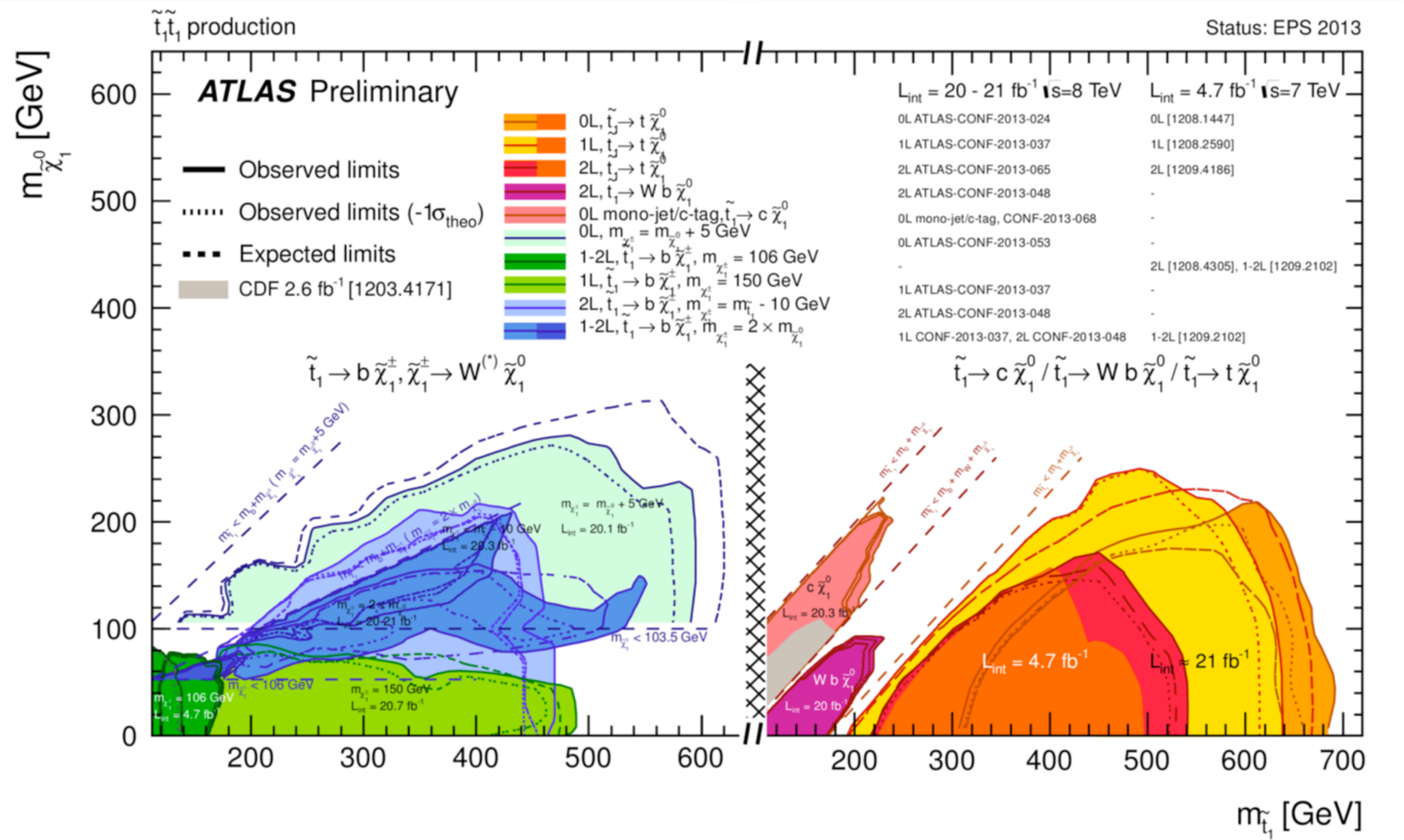}}
\caption{Summary of the dedicated ATLAS searches for stop pair production based on $20-21~\ifb$ of $pp$ collision data taken at $\sqrt{s} = 8~\tev$, and 4.7~\ifb\ of $pp$ collision data taken at $\sqrt{s} = 7~\tev$. From Ref.~\cite{summary-results}. \label{fg:natural}}
\end{figure}

If the gluino is too heavy to be produced at the LHC, direct $\t{t}_1\t{t}_1$ and $\t{b}_1\t{b}_1$ production is the only remaining possibility. If stop pairs are considered, two decay channels can be distinguished depending on the mass of the stop: $\t{t}_{1}\to b\t{\chi}_1^{\pm}$ and $\t{t}_{1}\to t\t{\chi}_1^0$. ATLAS has carried out a wide range of different analyses in each of these modes at both 7~\tev\ and 8~\tev\ centre-of-mass energy. In all these searches, the number of observed events has been found to be consistent with the SM expectation. Limits have been set on the mass of the scalar top for different assumptions on the mass hierarchy scalar top-chargino-lightest neutralino, as shown in the left (right) panel of Fig.~\ref{fg:natural} for $\t{t}_{1}\to b\t{\chi}_1^{\pm}$ ($\t{t}_{1}\to t\t{\chi}_1^0$) decays. 

For the former scenario, a scalar top quark of mass of up to 480~\gev\ is excluded at 95\% CL for a massless neutralino and a 150~\gev\ chargino. For a 300~\gev\ scalar top quark and a 290~\gev\ chargino, models with a neutralino with mass lower than 180~\gev\ are excluded at 95\% CL. For the case of a high-mass stop decaying to a top and a \X, analyses requiring one, two or three isolated leptons, jets and large \met\ have been carried out. Stop masses are excluded between 200~\gev\ and 680~\gev\ for massless neutralinos, and stop masses around 500~\gev\ are excluded along a line which approximately corresponds to neutralino masses up to 250~\gev. It is worth noting that a monojet analysis with $c$-tagging is deployed to cover part of the low-$m_{\t{t}_1}$, low-$m_{\t{\chi}_1^0}$ region through the $\t{t}_1\to c \t{\chi}_1^0$ channel~\cite{monojet}. 

\begin{figure}[htb]
{\includegraphics[width=0.6\textwidth]{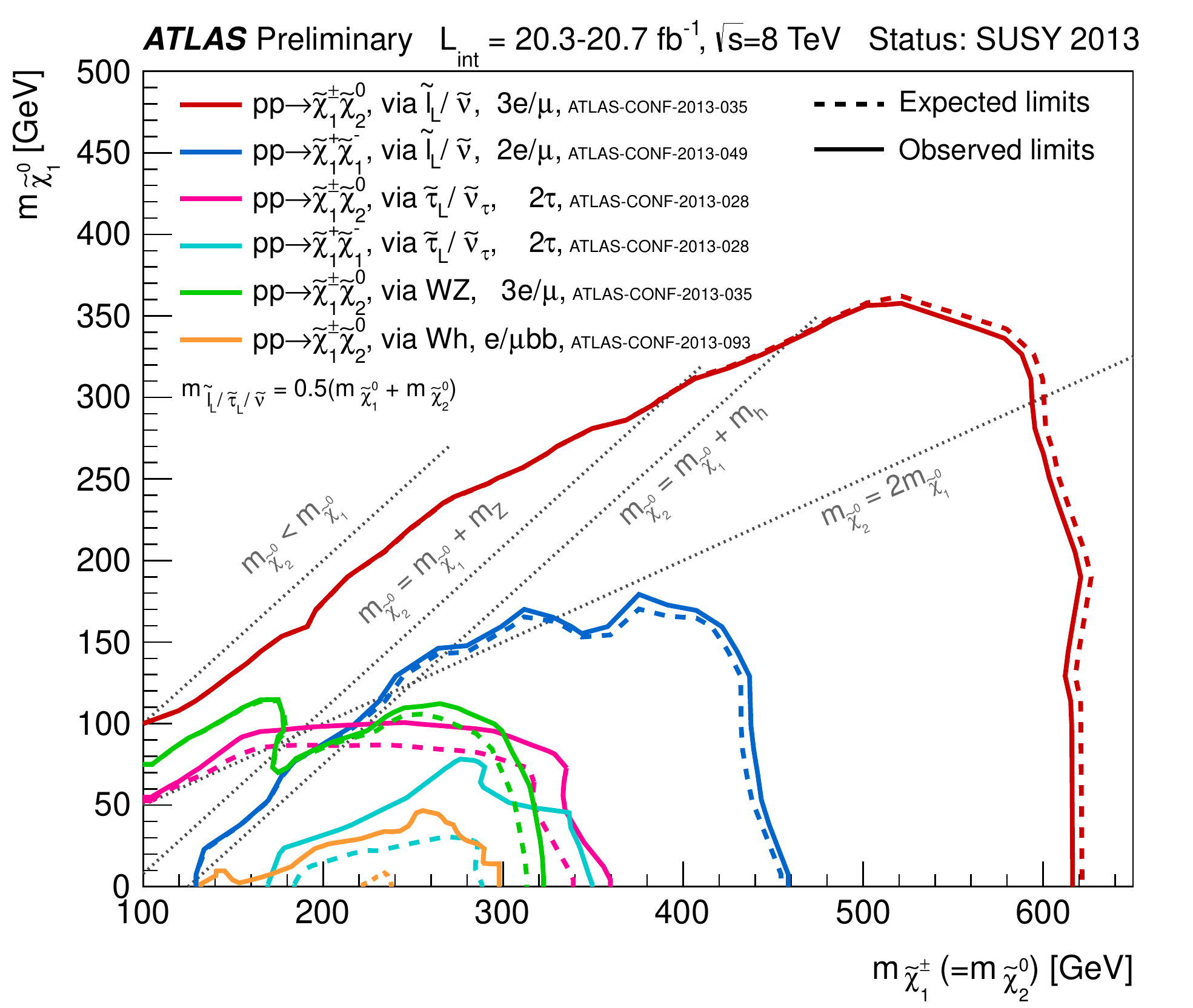}}
\caption{Summary of ATLAS searches for electroweak production of charginos and neutralinos based on 20~\ifb\ of $pp$ collision data at $\sqrt{s} = 8~\tev$. From Ref.~\cite{summary-results}.\label{fg:ewkino}}
\end{figure}

If all squarks and gluinos are above the TeV scale, weak gauginos with masses of few hundred GeV may be the only sparticles accessible at the LHC. As an example, at $\sqrt{s}Ê= 7~\tev$, the cross-section of the associated production $\t{\chi}_1^{\pm}\t{\chi}_2^0$ with degenerate masses of 200~\gev\ is above the 1-\tev\ gluino-gluino production cross section by one order of magnitude. Chargino pair production is searched for in events with two opposite-sign leptons and \met\ using a jet veto, through the decay $\t{\chi}_1^{\pm} \to \ell^{\pm}\nu\t{\chi}_1^0$. A summary of related analyses performed by ATLAS is shown in Fig.~\ref{fg:ewkino}.  Exclusion limits at 95\% confidence level are shown in the $(m_{\t{\chi}_0^{\pm}},\,m_{\X})$ plane.  The dashed and solid lines show the expected and observed limits, respectively, including all uncertainties except the theoretical signal cross section uncertainties. 

Charginos with masses between 140 and 560~\gev\ are excluded for a massless LSP in the chargino-pair production with an intermediate slepton/sneutrino between the $\t{\chi}_1^{\pm}$ and the $\t{\chi}_1^0$. If $\t{\chi}_1^{\pm}\t{\chi}_2^0$ production is assumed instead, the limits range from 11 to 760~\gev. The corresponding limits involving intermediate $W$, $Z$ and/or $H$ are significantly weaker.  

Searches are also performed in ATLAS for several signatures associated with the violation of $R$~parity (RPV).  As an example we highlight here the interpretation of null inclusive searches in the one-lepton channel~\cite{brpv} in the context of a model where RPV is induced through bilinear terms, depicted in the left panel of Fig.~\ref{fg:rpvll}. The results are obtained by combining the electron and muon channels. The band around the median expected limit shows the $\pm1\sigma$ variations on the median expected limit, including all uncertainties except theoretical uncertainties on the signal. The dotted lines around the observed limit indicate the sensitivity to $\pm1\sigma$ variations on these theoretical uncertainties. The thin solid black contours show the LSP lifetime. The result from the previous ATLAS search~\cite{brpv-prd} for this model is also shown. 

\begin{figure}[htb]
{\includegraphics[width=0.48\textwidth]{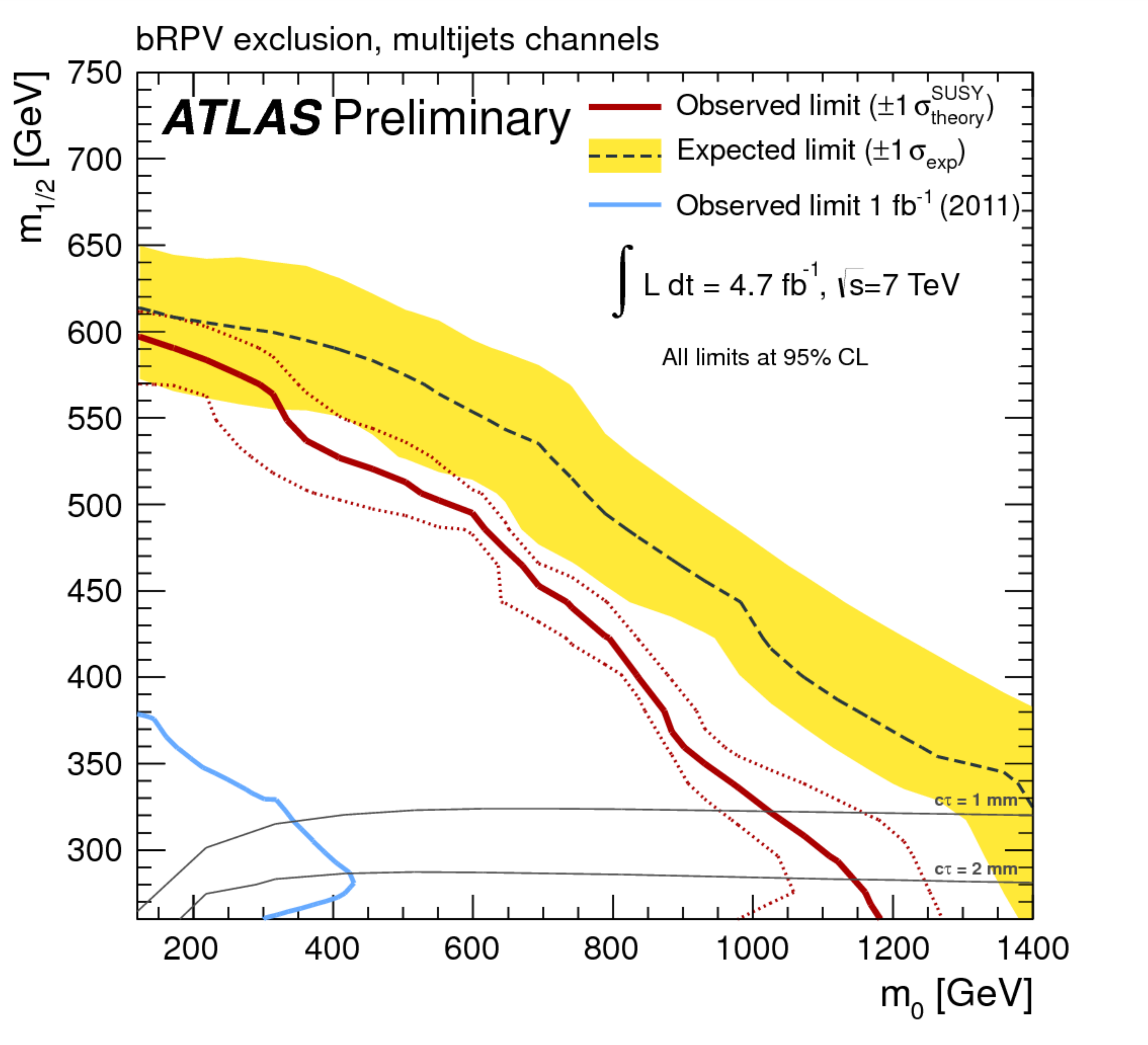}}
{\includegraphics[width=0.49\textwidth]{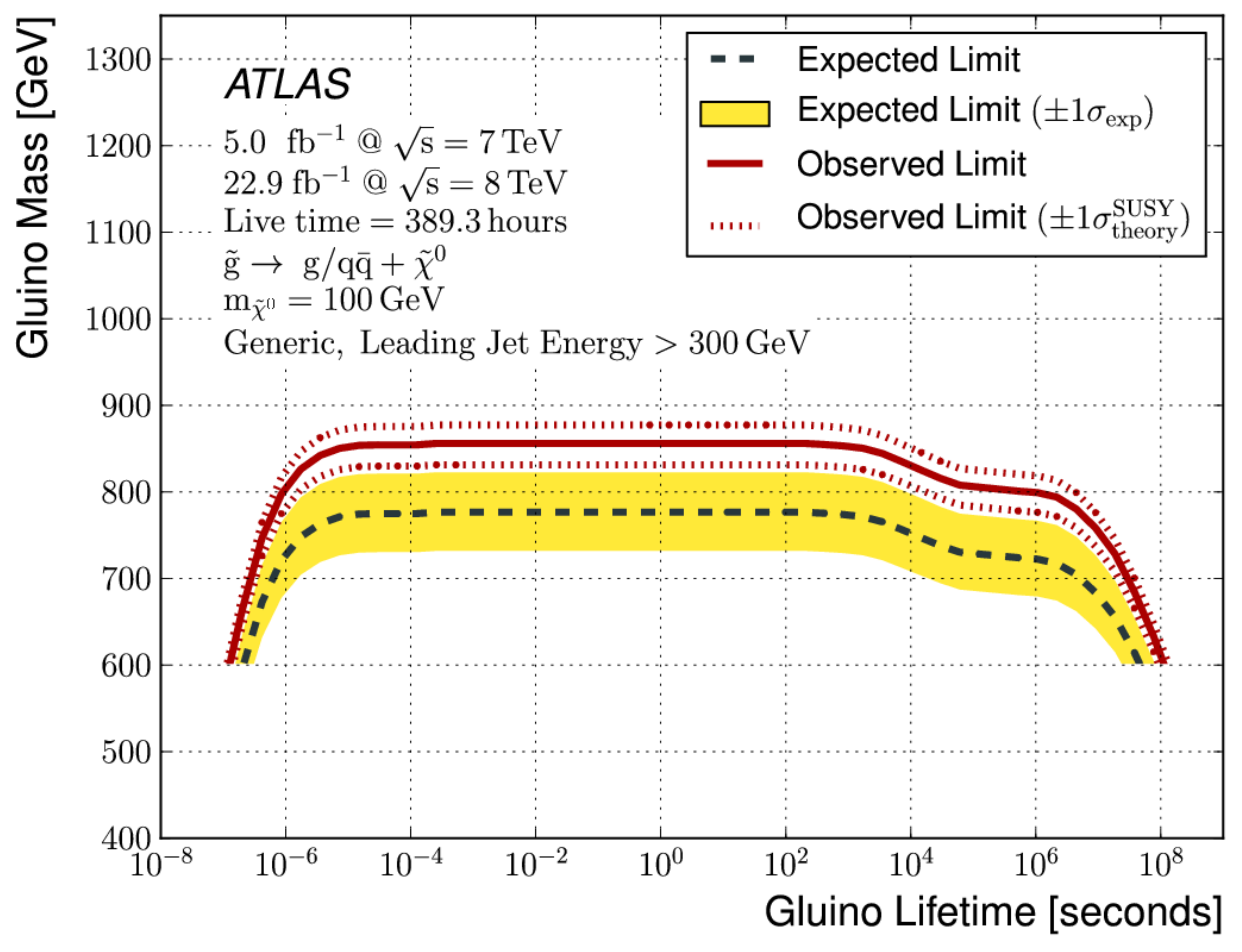}}
\caption{{\it Left:} Expected and observed 95\% CL exclusion limits in the bilinear $R$-parity violating model.  From Ref.~\cite{brpv}. {\it Right:} Bayesian lower limits on gluino mass versus its lifetime with R-hadron lifetimes in the plateau acceptance region between $10^{-5}$ and $10^3$ seconds. From Ref.~\cite{r-hadron}. \label{fg:rpvll}}
\end{figure}

In view of the null results in other SUSY searches, it became mandatory to fully explore the SUSY scenario predicting meta-stable or long-lived particles. These particles, not present in the Standard Model, would provide striking signatures in the detector and rely heavily on a detailed understanding of its performance. For instance, gluino, top squark, or bottom squark R-hadrons that have come to rest within the ATLAS calorimeter, and decay at some later time to hadronic jets and a neutralino have been searched for, using 5.0 and 22.9~\ifb\ of $pp$ collisions at 7~and 8~\tev, respectively. Selections based on jet shape and muon-system activity are applied to discriminate signal events from cosmic-ray and beam-halo muon backgrounds. In the absence of an excess of events, limits are set on gluino, stop, and sbottom masses for different decays, lifetimes, and neutralino masses. As shown in the right panel of Fig.~\ref{fg:rpvll}, with a neutralino of mass 100~\gev, the analysis excludes gluinos with mass below 832~\gev (with an expected lower limit of 731~\gev), for a gluino lifetime between $10~\mu{\rm s}$ and 1000~s in the generic R-hadron model with equal branching ratios for decays to $q\bar{q}+\X$ and $g+\X$. Under the same assumptions for the neutralino mass and squark lifetime, top squarks and bottom squarks in the Regge R-hadron model are excluded with masses below 379 and 344~\gev, respectively.

\section{Searches for exotic scenarios}\label{sc:exotics}

Various theoretical scenarios that attempt to short out some of the Standard Model problems, in addition to supersymmetry, have also been enquired by ATLAS. Several signature-driven analyses have been pursued:
\begin{compactitem}
\item resonances: dileptons, jets, photons, bosons, ...;
\item special particles: slow-moving, long-lived, ...;
\item \met\ plus other object(s);
\item other non-conventional signatures.
\end{compactitem}

Subsequently the outcome of these searches are interpreted in the context of specific models to obtain exclusion limits on masses, scales, etc: Extra-dimension scenarios, heavy gauge bosons, contact interactions, leptoquarks, new quarks, excited fermions, magnetic monopoles, new gauge bosons, etc. Given the wide range of possibilities, it is not possible to cover every analysis here, hence only some recent results are highlighted. 

A search for a dijet resonance with an invariant mass in the range between 130 and 300~\gev\ has been performed in the processes $pp\to W\text{jj}+X$ with $W\to\ell\nu$ and $pp\to Z\text{jj}+X$ with $Z\to\ell^+\ell^-$ $(\ell=e,\mu)$~\cite{lstc}. The data used correspond to 20.3~\ifb\ of $pp$ collision data recorded at $\sqrt{s}=8~\tev$ with ATLAS. The results are interpreted in terms of constraints on the Low Scale Technicolor (LSTC) model. No significant deviation from the Standard Model background prediction is observed. Upper limits on the production cross section times branching fraction are set for a hypothetical technipion $(\pi_{\rm T})$ produced in association with a $W$ or $Z$ boson from the decay of a technirho $(\rho_{\rm T})$ particle. The limits for the $W$ boson are shown in the left panel of Fig.~\ref{fg:reson}, assuming the mass relation of $m_{\rho_{\rm T}}=3/2\times m_{\pi_{\rm T}} + 55~\gev$. The inner and outer bands on the expected limit represent $\pm1\sigma$ and $\pm2\sigma$ variations, respectively. The LSTC predictions for the $\rho_{\rm T}$ cross section are also shown in blue. 

\begin{figure}[htb]
{\includegraphics[width=0.5\textwidth]{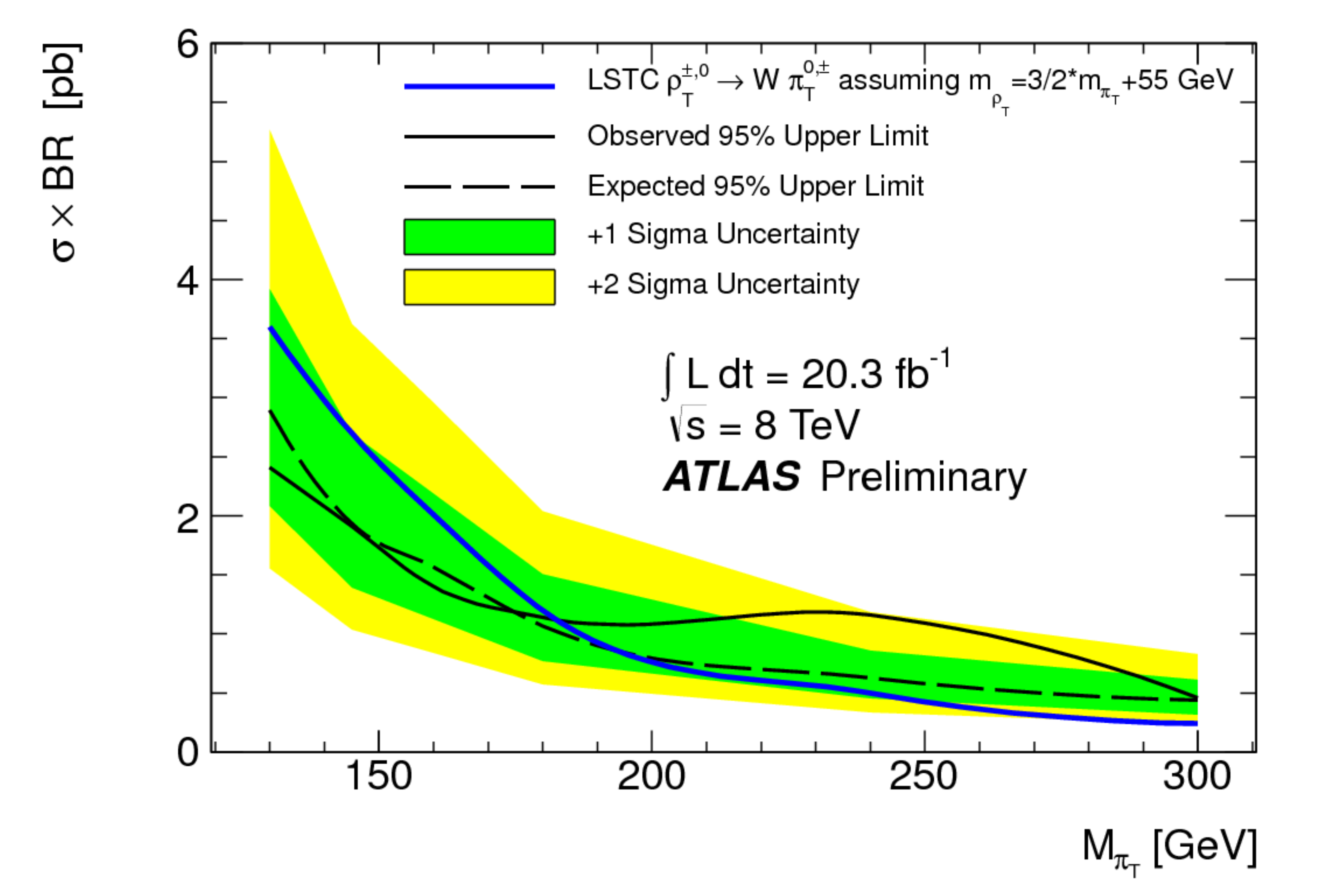}}
{\includegraphics[width=0.5\textwidth]{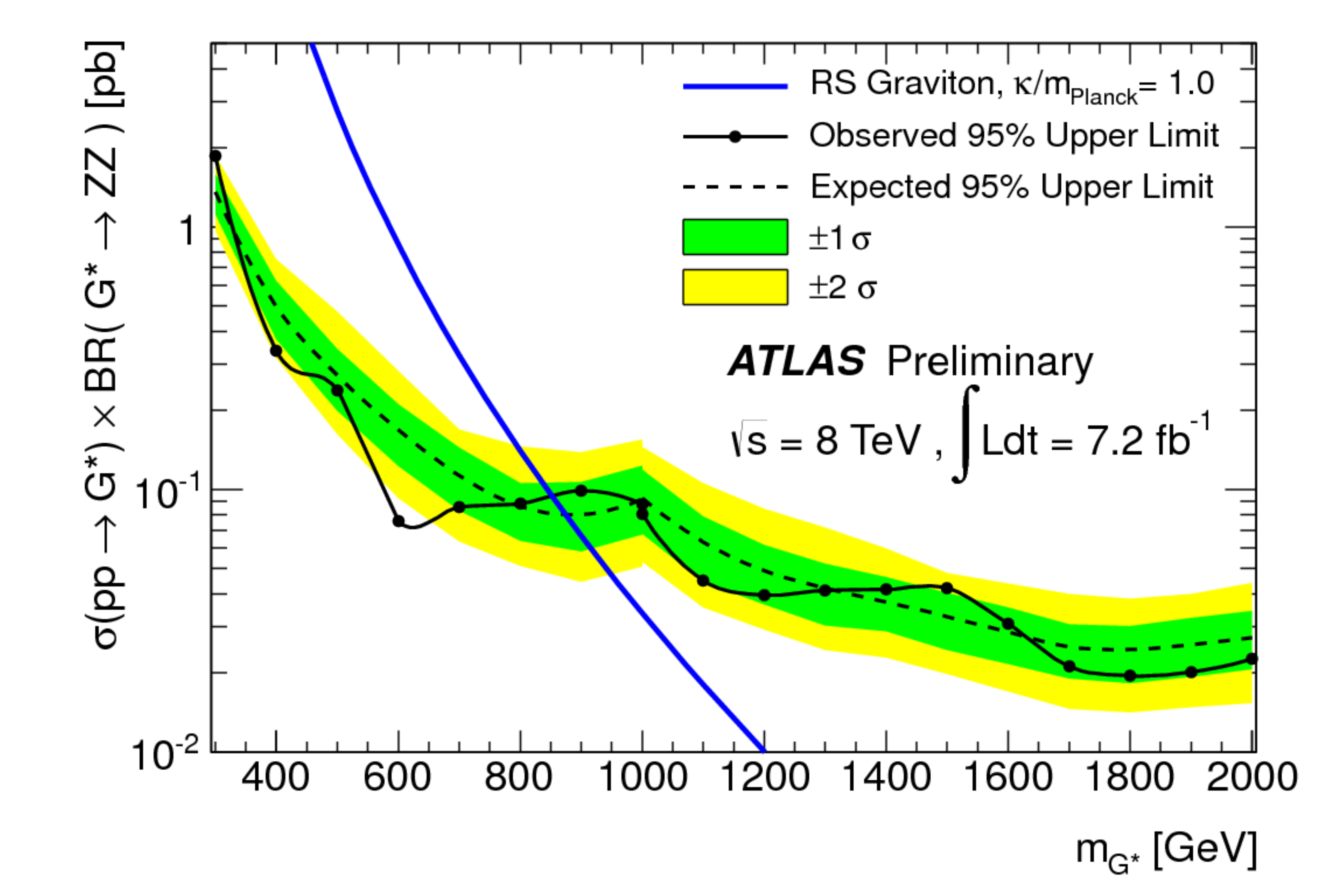}}
\caption{{\it Left:} Observed and expected 95\% CL upper limits on the $W\pi_{\rm T}$ cross section as a function of the mass of the technipion in the $W$jj channel. From Ref.~\cite{lstc}. {\it Right:} Observed and expected 95\% credibility level upper limits on $\sigma(pp\to G^*)\times BR(G^*\to ZZ)$ for the bulk RS graviton.  From Ref.~\cite{llqq}. \label{fg:reson}}
\end{figure}

A search for a heavy particle that decays to a pair of $Z$ bosons has been carried out using events recorded by ATLAS in $pp$ collisions at $\sqrt{s} = 8~\tev$ corresponding to an integrated luminosity of 7.2~\ifb. The analysis considers the $\ell\ell qq$ final state and uses the diboson mass reconstructed from the leptons and jets as the discriminating variable. No significant deviation in the mass distribution from a smoothly falling background distribution is observed. The upper limits set on the production cross section times branching fraction into a $ZZ$ boson pair for the bulk Randall-Sundrum (RS) graviton with coupling parameter of $\kappa/\bar{m}_{\text{Pl}} = 1.0$ are shown in the right panel of Fig.~\ref{fg:reson}. The leading-order theoretical prediction for the bulk RS model is also shown. The inner and outer bands on the expected limit represent $\pm1\sigma$ and $\pm2\sigma$ variations, respectively. The corresponding 95\% credibility level observed (expected) lower limit on the mass for the graviton is 850 (870)~\gev.

Weakly-interacting massive particles (WIMPs, $\chi$), stable neutral states, which exist in many extensions of the SM, provide a good candidate to explain the cosmological dark matter. The leading generic diagrams responsible for DM production at hadron colliders~\cite{dm-review} involve the WIMP pair-production plus the initial- or final-state radiation (ISR/FSR) of a gluon, photon, weak gauge boson $Z,\,W$ or a Higgs boson. The ISR/FSR particle is necessary to balance the two WIMPs' momentum, so that they are not produced back-to-back resulting in negligible \met. Therefore the search is based on selecting events high-\met\ events, due to the WIMPs, and a single jet~\cite{monojet2}, photon~\cite{monophoton} or boson~\cite{monowz} candidate. 

In the context of $pp\to\x\bar{\x}+W/Z$ production, ATLAS has searched for the production of $W$ or $Z$ bosons decaying hadronically and reconstructed as a single massive jet in association with large \met\ from the undetected $\chi\bar{\chi}$ particles~\cite{monowz}. For this analysis, the jet candidates are reconstructed using a filtering procedure referred to as large-radius jets. The 90\% limits on $\chi$-nucleon cross sections for spin-independent scattering are shown in Fig.~\ref{fg:mono} (left). They are compared to previous limits set by direct dark matter detection experiments and by the ATLAS 7~\tev\ monojet analysis~\cite{monojet2}. For the spin-independent case with the opposite-sign up-type and down-type couplings, the limits are improved by about three orders of magnitude. 

\begin{figure}[htb]
{\includegraphics[width=0.5\textwidth]{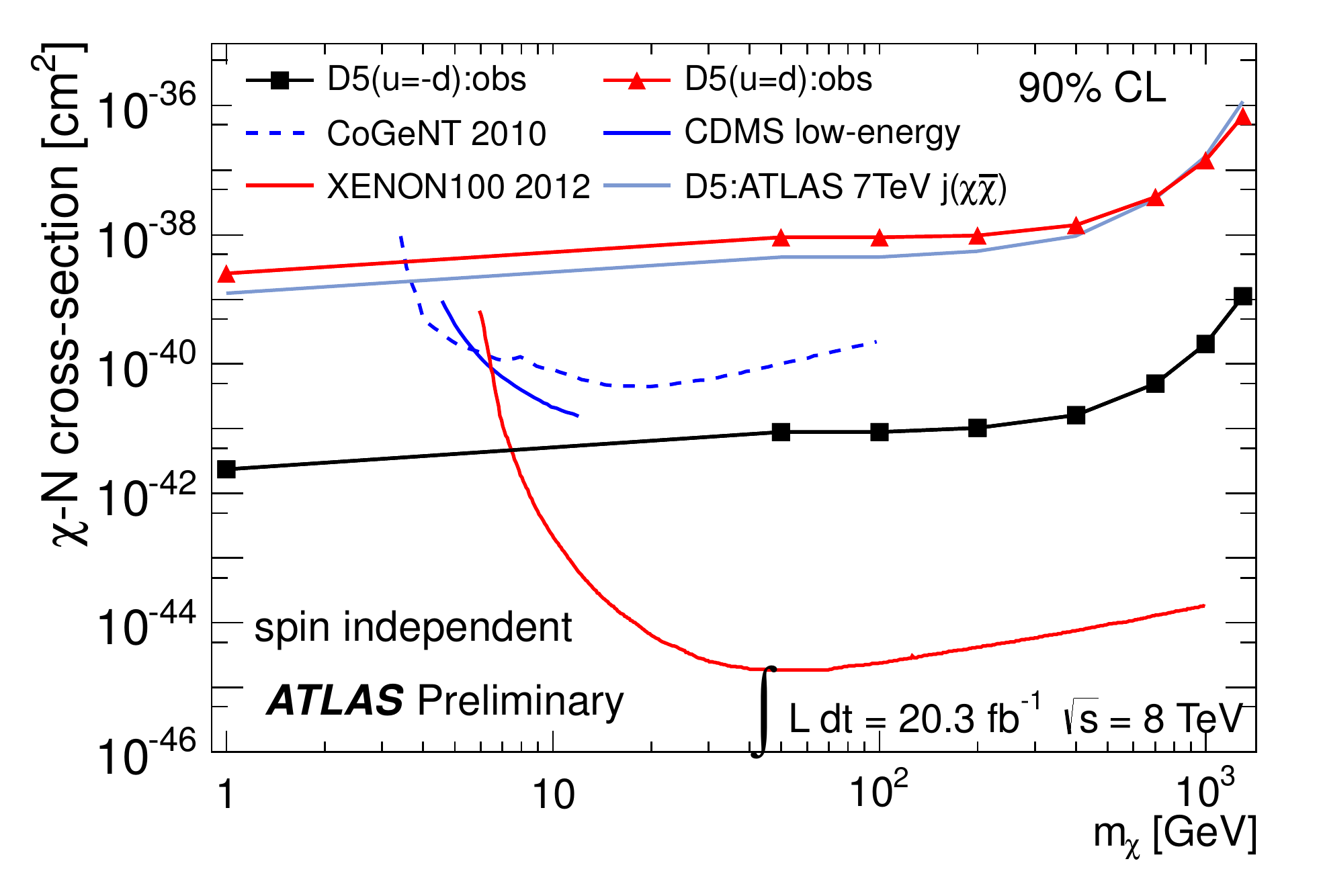}}
{\includegraphics[width=0.5\textwidth]{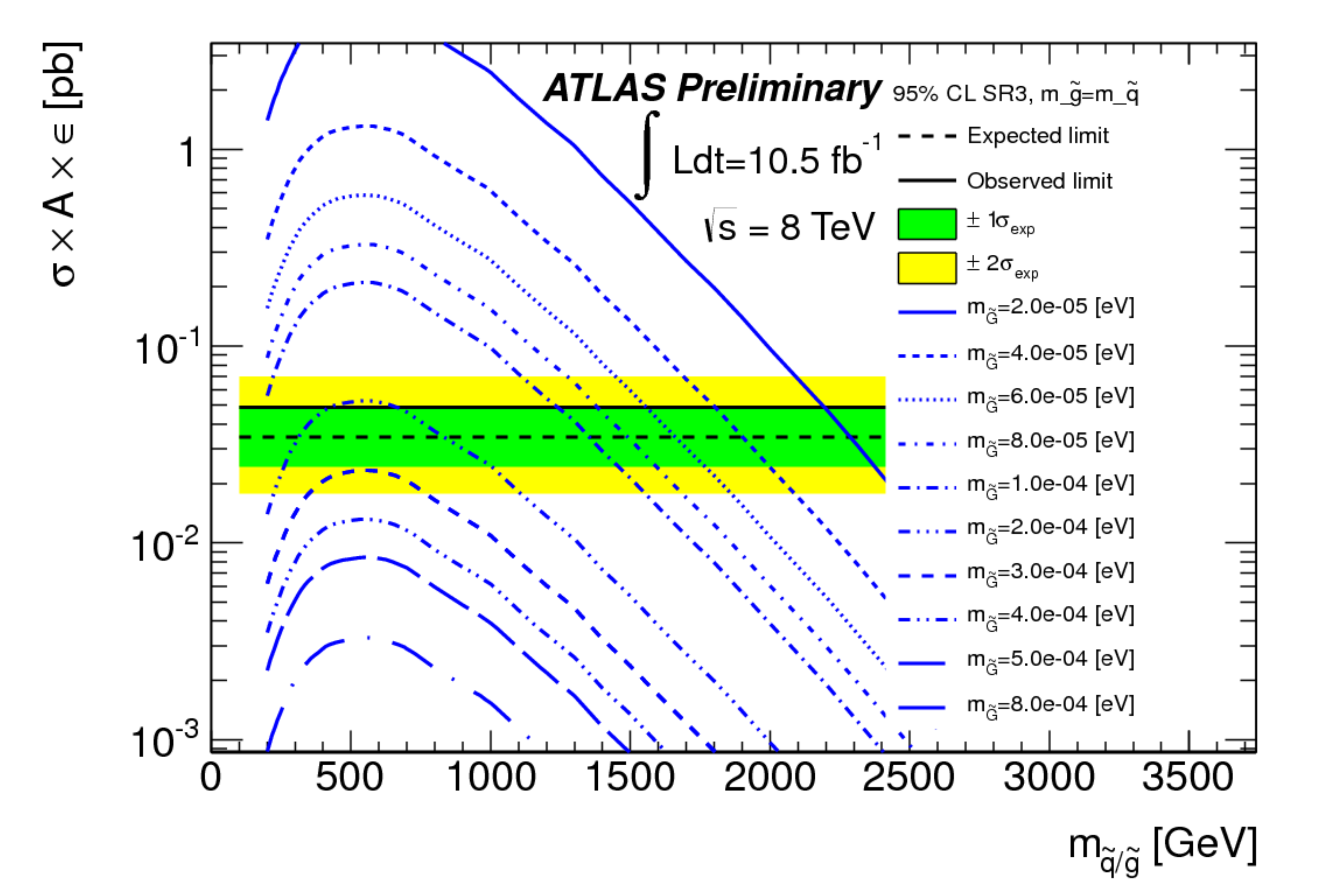}}
\caption{{\it Left:} 90\% limits on $\chi$-nucleon cross sections for spin-independent scattering obtained by the mono-$W/Z$ analysis. From Ref.~\cite{monowz}. {\it Right:} Cross section times acceptance times efficiency for the gravitino plus $\t{q}/\t{g}$ production as a function of the $\t{q}/\t{g}$ mass for different values of the gravitino mass. From Ref.~\cite{monojet2}. \label{fg:mono}}
\end{figure}

This search is sensitive to WIMP pair production, as well as to other DM-related models, such as invisible Higgs boson decays ($WH$ or $ZH$ production with $H\to\chi\bar{\chi}$). Moreover the monojet analysis~\cite{monojet2} is pertinent for exploring the production of light grativinos in association with gluinos or squarks in a gauge-mediated supersymmetric model. The cross section times acceptance times efficiency for the gravitino plus $\t{q}/\t{g}$ production as a function of the $\t{q}/\t{g}$ mass in the case of mass degenerate squark and gluinos is shown in Fig.~\ref{fg:mono} (right). The comparison of these curves with the acquired model-independent upper limit from the monojet search leads to the best lower bound to date on the gravitino mass. Exclusion limits on models with large extra spatial dimensions and on pair production of weakly interacting dark matter candidates have been obtained as well.

The leading quadratically-divergent contribution to the Higgs mass is the top quark loop, making particles closely coupled to the top quark natural candidates for relatively light manifestations of new physics. Vector-like top partners, i.e.\ fermions for which the right- and left-chiral components follow the same transformation rules, present a rich phenomenology. For heavy top-like partners, new decay modes open up: $T\to Wb$, $T\to Ht$ and $T\to Zt$ are possible. In addition, such partners can be part of multiplets that can include for example a charge $+5/3$ quark $(T_{+5/3})$, a heavy $B$ quark or a charge $-4/3$ quark $(Y_{-4/3})$. This can lead to a rich set of signatures which generally include many third-generation particles in the final state.

In the case of pair production of $T_{5/3}$, the decay chain $T_{+5/3}\to W^+t\to W^+W^+b$ can lead to a pair of same-sign leptons, if both $W$ bosons decay leptonically~\cite{vlq1}. In the charge $2/3$ case, for pair production the $T\to Wb$ decay mode leads to the same signature as SM $t\bar{t}$ production, so that a good discriminant is needed, such as the reconstructed heavy quark mass~\cite{vlq2}. The $T\to Ht$ decay mode leads to an abundance of $b$-jets, since the final state consists of $WbbbX$, where $X$ represents the decay products of the second $T$ quark, and contains at least one $b$ quark~\cite{vlq3}. This allows the application of hard selection cuts in this channel, requiring at least six jets of which at least four should be $b$-tagged. For the third decay mode, $T\to Zt$, requiring a leptonically-decaying $Z$ boson offers excellent background suppression. After imposing a high \pt\ cut on the $Z$, ATLAS uses the reconstructed $Zb$ mass (using the leading $b$-jet) as a discriminating variable~\cite{vlq4}. This search is also sensitive to production of a heavy $B$ quark decaying to $Zb$, for which this discriminating variable is even more powerful. Figure~\ref{fg:vlq-tt} summarises the limits obtained by the different ATLAS $T$~quark searches.

\begin{figure}[htb]
{\includegraphics[width=\textwidth]{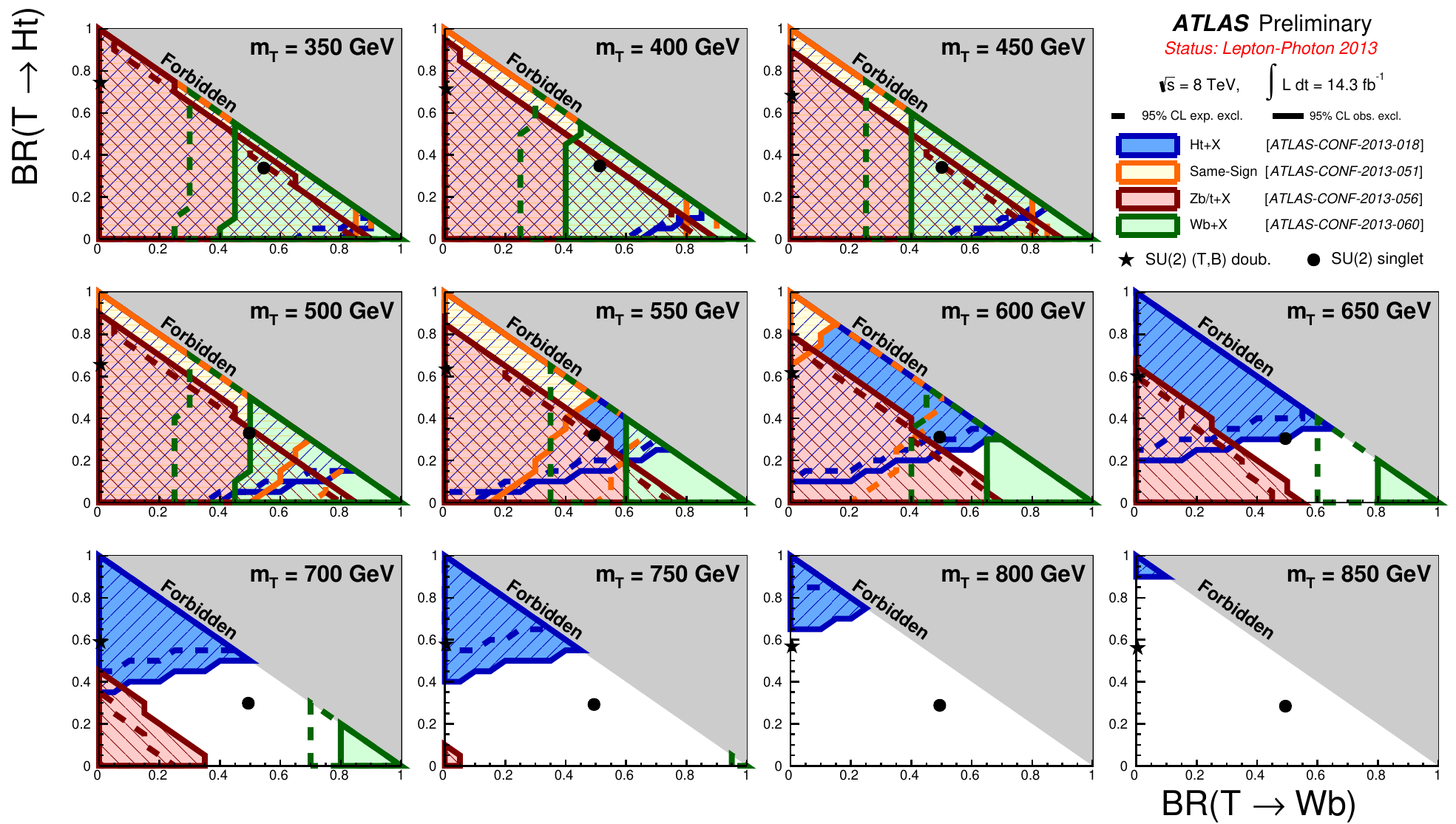}}
\caption{Full summary of ATLAS searches for vector-like $T$ quark with 14~\ifb\ of 8~\tev\ data. Excluded regions are drawn sequentially for each of the analyses in chronological order and overlaid (rather than combined) in each of the figures.  From Ref.~\cite{summary-results}. \label{fg:vlq-tt}}
\end{figure}

\section{Summary and outlook}\label{sc:summary}

Excellent performance from ATLAS experiment and the LHC has been observed during the 7~\tev\ and 8~\tev\ runs in 2011-2012. The first results from Run~I include high precision in measurements of SM processes, the discovery of a boson consistent with the SM Higgs and improved limits on BSM scenarios. The analysis of the latest data is still on-going and more exciting results are expected to be released soon. The Particle Physics community is looking forward to LHC resuming collisions at $13-14~\tev$ after the 2013-2014 LHC long shutdown.

\section*{Acknowledgments}

The author acknowledges support by the Spanish Ministry of Economy and Competitiveness (MINECO) under the projects FPA2009-13234-C04-01 and FPA2012-39055-C02-01, by the Generalitat Valenciana through the project PROMETEO~II/2013-017 and by the Spanish National Research Council (CSIC) under the JAE-Doc program co-funded by the European Social Fund (ESF). 


\end{document}